\documentclass[journal,12pt, draftclsnofoot, onecolumn]{IEEEtran}
\usepackage{graphicx}
\usepackage{amsmath}
\usepackage{subfig}
\usepackage{mathtools}
\usepackage{epsfig}
\usepackage{float}
\usepackage{lipsum}
\usepackage{stfloats}
\usepackage{array}
\usepackage{amssymb}
\usepackage{cite}
\usepackage{url}
\usepackage{amsthm}
\usepackage{algorithm2e}
\usepackage{algorithmic}
\usepackage{multirow}
\usepackage{upgreek}
\usepackage{diagbox}
\usepackage{dsfont}

\normalsize
\newcommand{\spazio}{\rule[-6mm]{0mm}{0mm}}

\newcommand{\Phib}{\Phi_{\rm b}}
\newcommand{\lambdab}{\lambda_{\rm b}}
\newcommand{\Phiu}{\Phi_{\rm u}}
\newcommand{\lambdau}{\lambda_{\rm u}}

\newcommand{\Cexact}{C_{\sf exact}}
\newcommand{\Cub}{C_{\sf ub}}
\newcommand{\barCexact}{\bar{C}_{\sf exact}}
\newcommand{\barCub}{\bar{C}_{\sf ub}}

\newcommand{\shadSD}{\upsigma_{\scriptscriptstyle \rm dB}}

\newcommand{\mathbE}{\mathbb{E}}
\newcommand{\mathbP}{\mathbb{P}}
\newcommand{\mathbR}{\mathbb{R}}

\newcommand{\rk}{{r}_k}

\newcommand{\ro}{{r}_0}
\newcommand{\Hk}{H_k}
\newcommand{\Ho}{H_0}

\newcommand{\sk}{s_k}
\newcommand{\so}{s_0}
\newcommand{\zo}{z}

\newcommand{\SIRo}{\mathsf{SIR}}
\newcommand{\SINRo}{\mathsf{SINR}}

\newcommand{\Cbar}{\bar{C}}
\newcommand{\E}{\mathbb{E}}
\newcommand{\erf}{\mathrm{erf}}

\newcommand{\Nt}{N_{\rm t}}
\newcommand{\Nr}{N_{\rm r}}

\newcommand{\sinc}{\mathrm{sinc}}

\newcommand{\Cmimo}{C^{\scriptscriptstyle \mathsf{MIMO}}_{\scriptscriptstyle 2\times2}}

\newcommand{\Gfbr}{Q}

\theoremstyle{definition}

\newtheorem{exmp}{Example}

\newlength{\figwidth}
\begin{document}

\title{Ergodic Spectral Efficiency in MIMO \\ Cellular Networks}

\author{Geordie~George, Ratheesh~K.~Mungara, Angel~Lozano, and Martin Haenggi
\thanks{G. George, R. K. Mungara and A. Lozano are with the Department of Information and Communication Technologies, Universitat Pompeu Fabra (UPF), 08018 Barcelona, Spain. E-mail: \{geordie.george, ratheesh.mungara, angel.lozano\}@upf.edu.

Martin Haenggi is with the University of Notre Dame, Notre Dame, IN 46556, USA. E-mail: mhaenggi@nd.edu.

This work was supported by Project TEC2015-66228-P (MINECO/FEDER, UE), by the European Research Council under the H2020 Framework Programme/ERC grant agreement 694974, and by the U.S.~NSF through award CCF 1525904. This paper is accepted for presentation in part at the 2017 IEEE Int'l Conference on Communications (ICC). 
}
}


\maketitle

\begin{abstract}
This paper shows how the application of stochastic geometry to the analysis of wireless networks is greatly facilitated by (\emph{i}) a clear separation of time scales, (\emph{ii}) the abstraction of small-scale effects via ergodicity,
and (\emph{iii}) an interference model that reflects the receiver's lack of knowledge of how each individual interference term is faded. These procedures render the analysis both more manageable and more precise, as well as more amenable to the incorporation of subsequent features. In particular, the paper presents analytical characterizations of the ergodic spectral efficiency of cellular networks with single-user multiple-input multiple-output (MIMO) and sectorization. These characterizations, in the form of easy-to-evaluate expressions, encompass the coverage, the distribution of spectral efficiency over the network locations, and the average thereof.
\end{abstract}


\section{Introduction}
\label{intro}

\IEEEPARstart{S}{tochastic} geometry is quickly becoming an indispensable instrument in wireless network analysis.
By mapping the empirical distribution of transmitter and receiver locations to appropriate point processes, it becomes possible to apply a powerful and expanding toolkit of mathematical results. This offers a complement, and increasingly even an outright alternative, to the Monte-Carlo simulations that have long been the workhorse of wireless network design.

Although a stochastic modelling of transmitter and receiver locations may seem mostly amenable to ad hoc networks, which are devoid of fixed infrastructure, a seminal paper by Andrews \emph{et al.} \cite{Andrews-Baccelli-Ganti-coverage} proved the truly remarkable effectiveness of stochastic modelling also in cellular networks---even with simple Poisson point processes (PPPs).
Indeed, while it may appear that more sophisticated spatial distributions could better capture the relative regularity of actual base station (BS) placements, because of shadowing it is the case that PPPs lead to remarkably precise characterizations of signal strengths and interference, and thus of all ensuing performance measures.
In fact, as shown in \cite{7081363,keeler2014wireless} and expounded later in this paper, PPP-based characterizations represent the limit to which actual behaviors converge as the shadowing strengthens.

Altogether, the irruption of stochastic geometry is a transcendent development in wireless research, and it is reasonable to expect its importance to grow even further as networks become denser and more heterogeneous \cite{What5G,Boccardi-5G}.
Important contributions to the advancement of the discipline in the context of wireless networks include
\cite{StochGeoIntro-Haenggi-JSAC,Poisson-field-partI,Poisson-field-partII,MHaenggi12,MHaenggi-GinibrePP14,7104127,HaenggiGC2015,5621983,Dhillon-HetNet-JSAC12, Mukherjee-HetNet-JSAC12, Offloading-Andrews-TWC,ELSawy13,  Poisson-Cognitive-Haenggi,7243344,net:Guo14icc,7448861,di2015stochastic,dhillon2013downlink,MIMO-reTX-stochgeo,MUMIMO-Li-TCOM16,MIMO-TRXDiv-stochgeo,MIMO-noncoh-coop}.


The present paper deals with the ergodic spectral efficiency of Poisson cellular networks, a quantity already tackled in works dating back to \cite{Andrews-Baccelli-Ganti-coverage}. Our analysis, however, relies on a different modeling approach for the interference. This takes us on a different route, one that proves greatly advantageous because it yields solutions that are both simpler and more precise than previous ones, and, most importantly, because it opens the door to accommodating key ingredients---such as MIMO and cell sectorization---that seemed previously elusive.
The modeling approach that unlocks these new analytical possibilities 
 is not arbitrary, but rather based on sound arguments:
\begin{enumerate}
\item A clean separation between small- and large-scale channel features, in recognition that the phenomena that give rise to these features are distinct.
\item An unwavering embrace of ergodic performance metrics with respect to the small-scale features, in recognition that such small-scale ergodicity reflects well the operating conditions of modern wireless systems \cite{5374062,Loz-Jin12}.
\item The admission that each receiver can track the fading of its intended signal, but not the fading of each individual interference term.
\end{enumerate}

With small- and large-scale features decoupled, ergodicity enables abstracting out the former so as to
focus the stochastic geometry analysis where it makes a difference (on the large-scale aspects), reaping the most out of its potent machinery. As mentioned, this allows advancing the analysis on all fronts: tractability, accuracy, and generality. In particular, and to the extent of our knowledge, the spectral efficiency expressions obtained in this paper are the first such characterizations that incorporate MIMO spatial multiplexing. Likewise, sectorization is also readily included.

Besides providing new expressions for quantities of interest, the present paper seeks to promote the importance of keeping the two foregoing arguments present when applying stochastic geometry to wireless networks.
For conceptual clarity, these arguments are herein elaborated on the basis of a cellular network where each user is served by a single BS, and where only the downlink is considered. However, the arguments apply equally to networks featuring BS cooperation, and to the uplink, only with certain expressions suitably replaced by generalizations or counterparts.

\section{Network Modelling}
\label{sys_model}

\subsection{Separation of Scales}

Rooted in extensive propagation measurements, the separation between large-scale propagation phenomena (distance-dependent path loss and shadowing) and small-scale multipath fading has been instrumental in the study of wireless networks since their onset, greatly facilitating characterizations that would otherwise be unwieldy \cite{jakes1994microwave}. The premise of this separation is that, over suitably small distances (tens to hundreds of wavelengths), the large-scale phenomena remain essentially unchanged and only small-scale variations transpire.
This allows delineating local neighborhoods around transmitters and receivers wherein the small-scale channel behavior conforms to a stationary random process whose distribution is dictated by the large-scale features.  Then, through the user velocity, this space-domain random process maps to a time-domain process. Moreover, under the very mild condition that the Doppler spectrum be free of delta functions, this time-domain process is ergodic.

For frequencies and velocities in the widest possible range of interest, the dwell time in a local neighborhood is far longer than the extension of signal codewords. Thus, large-scale features can be regarded as constant over an individual codeword. 
Alternatively, the small-scale fading may or may not remain constant over a codeword, depending on the coherence of such fading and on how codewords are arranged in time and frequency, and this dichotomy gives rise to two classic information-theoretic idealizations of the channel over the horizon of a codeword:
\begin{itemize}
\item \emph{Nonergodic}. Fading random, but fixed over the codeword. 
\item \emph{Ergodic}. Fading random and exhibiting sufficiently many values over the codeword to essentially reveal its distribution.
\end{itemize}
 
These two idealizations, in turn, map respectively to outage and ergodic definitions for the spectral efficiency. 
Although both are useful, the ergodic definition is the most representative in modern systems where codewords can be interspersed over very wide bandwidths, across hybrid-ARQ repetitions, and possibly over multiple antennas, and they can be subject to scheduling and link adaptation. As argued in \cite{5374062,Loz-Jin12}, the balance of these mechanisms is indeed best abstracted by ergodic spectral efficiencies involving expectations over the small-scale fading, with the large-scale features held fixed. It is at this point that stochastic geometry should enter the analysis, when the small-scale effects have been abstracted out and we can zoom out to cleanly examine the large-scale ones.

\subsection{Large-scale Modeling}
\label{GoRunning}

We consider the downlink of a cellular network, initially with omnidirectional antennas (to be generalized to sectorized antennas in Section \ref{sectorization}), where
the signals are subject to path loss with exponent $\eta > 2$ and shadowing.


Suppose that the BS positions are agnostic to the radio propagation.
It has been recently shown \cite{7081363,keeler2014wireless} that, regardless of what those BS positions are (under only a very mild homogeneity condition), an increasing shadowing standard deviation $\shadSD$ renders the network progressively PPP-like from the vantage of any given user, i.e., it makes the powers that a user receives from any population of BSs look as if they originated from PPP-distributed BSs. This important observation, upheld even if the shadowing is correlated \cite{NRoss2016},
strongly justifies the modelling assumption of PPP-located BSs.
Ironically then, shadowing, a nuisance in the study of regular geometries, simplifies the stochastic modelling of networks by making them all look alike propagation-wise regardless of their underlying geometry.
And, although the convergence to a PPP behavior is asymptotic in $\shadSD$, values of interest suffice for networks to look essentially Poissonian. In particular, it is shown in \cite{7081363} that, for a regular lattice of BSs spawning hexagonal cells and $\eta=4$, $\shadSD = 12$ dB suffices to render the received powers indistinguishable (with $99\%$ confidence in a Kolmogorov-Smirnov test) from those in a Poisson network, $90\%$ of the time.
In Section \ref{hexagons} we provide further evidence supporting the suitability of a PPP model for the analysis of lattice networks with relevant values of $\shadSD$.

The foregoing convergence is a powerful argument in favor of a PPP model for the spatial distribution of BSs, say a process $\Phib \subset \mathbR^2$, without the need for explicit modeling of the shadowing as it is already implicitly captured by the Poisson nature of the network. The density of $\Phib$, say $\lambdab$, depends on the type and strength of the shadowing in addition to the actual positions of the BSs \cite{4675736,ShadowingPPP,5678764,zhang2014stochastic}.

Turning now to the spatial distribution of users, a good starting point is to model it as an independent PPP $\Phiu$ with density $\lambdau$. (This model could be refined to incorporate user clustering tendencies as well as dependences between the positions of users and BSs \cite{guo2013spatial,deng2015heterogeneous,Dhillon15,saha2016enriched}.)

Without loss of generality, the analysis can be conducted from the perspective of a user at the origin, which becomes the typical user under expectation over $\Phib$.
Denoting by $\rk$ the distance between such user and the $k$th BS, whose location---recall---is distributed according to $\Phib$, we index the BSs in increasing order of $\rk$, i.e., such that $\rk < {r}_{k+1}$ for $k \in \mathbb{N}_0$. Since, in terms of $\Phib$, the only large-scale propagation mechanism at play is path loss, the user at the origin receives the strongest power from the BS at $\ro$, which we deem the serving BS.


\subsection{Small-scale Modeling}

Let the communication be SISO for now, i.e., with BSs and users having a single antenna, and further let each receiver be privy to the fading of only its intended signal.
Denoting by $P$ the power measured at 1 m from a BS transmitter, at symbol $n$ the user at the origin observes
\begin{align}
\label{eq:rec_y}
y[n] = \sqrt{P \, \ro^{-\eta}} \, \Ho[n] \, \so[n] + \zo[n] ,
\end{align}
where the leading term is the intended signal from the serving BS while
\begin{align}
\label{eq:aggr_z}
\zo[n] = \sum\limits_{k=1}^{\infty} \sqrt{P \, \rk^{-\eta}} \, \Hk[n] \, \sk[n] + v[n]
\end{align}
is the aggregate interference from all other BSs, plus thermal noise $v$. 
In turn, $\sk$ is the signal transmitted by the $k$th BS and $\Hk$ is the associated fading coefficient.

The fading coefficients $\{ H_{k} \}_{k=0}^\infty$ are independent and of unit power, but they are otherwise arbitrarily distributed, whereas $v \sim \mathcal{N}_{\mathbb{C}}(0,N_0)$. The signal is $\sk \sim \mathcal{N}_{\mathbb{C}}(0,1)$, a choice that is justified later. 


Conditioned on $\{ \rk\}_{k=0}^\infty$, which are fixed as far as the small-scale modeling is concerned, the instantaneous SINR
at symbol $n$ is
\begin{align}
\label{eq:SIR_wrong}
\SINRo[n] &= \frac{P \, \ro^{-\eta} \, |\Ho[n]|^2 }{P \sum_{k=1}^{\infty} \rk^{-\eta} \, |\Hk[n]|^2 + N_0} .
\end{align}

\section{Interference Modeling}

With $\Ho[1],\ldots,\Ho[N]$ known at the receiver, the mutual information (in bits/symbol) over codewords spanning $N$ symbols is
\begin{align}\nonumber
\frac{1}{N} I \Big( \! s_0[1],\ldots,s_0[N] ; y[1],\ldots,y[N] 
\big| \Ho[1],\ldots,\Ho[N] , \{ \rk \} \! \Big)
\end{align}
which, with IID codeword symbols, becomes
\begin{equation}
\label{Johan}
\frac{1}{N} \sum_{n=1}^N I \big( s_0[n] ; y[n] \, \big| \Ho[n] , \{ \rk \}_{k=0}^\infty \big) .
\end{equation}
As argued earlier, codewords are nowadays long enough---thousands of symbols---and arranged in such a way---interspersed in time, frequency, and increasingly across antennas---so as
to experience sufficiently many fading swings for an effective averaging of the mutual information to take place over the small-scale fading. 
From the stationarity and ergodicity of the small-scale fading over the codeword, the averaging in (\ref{Johan}) becomes an expectation over $\Ho$ and confers the significance of the ergodic spectral efficiency (in bits/s/Hz)
\begin{align}
\label{marathon}
\!\!\!\Cexact & = \mathbb{E}_{\Ho} \Big[ I \big( s_0 ; y \, \big| \Ho , \{ \rk \}_{k=0}^\infty \big) \Big] \\
& = \mathbb{E}_{\Ho} \! \left[ I \! \left(\! s_0 ; \sqrt{P \, \ro^{-\eta}} \, \Ho \, \so + z \, \Big| \Ho , \{ \rk \}_{k=0}^\infty \! \right) \right] .
\end{align}
This quantity, a baseline in the sequel, does not admit explicit expressions. Rather, the evaluation of $\Cexact$ requires computationally very intensive Monte-Carlo simulations (cf. App.~\ref{App:prelims})
and a $64$-core high-performance computing cluster is employed to generate the corresponding results throughout the paper; for all such results, 99\% confidence intervals are given.

Let us examine the local distribution, for some given $\{ \rk \}_{k=1}^\infty$, of the interference-plus-noise $z$ as defined in (\ref{eq:aggr_z}). The first thing to note is that, without further conditioning on $\{ H_k \}_{k=1}^\infty$, i.e., without the receiver knowing the fading coefficients from \emph{all} interfering BSs, the distribution of $z$ over the local neighborhood is generally not Gaussian. Conditioned only on $\{ \rk \}_{k=1}^\infty$, the distribution of $z$ is actually highly involved; in Rayleigh fading, for instance, it involves products of Gaussians.
While the non-Gaussianity of $z$ is irrelevant to the SINR, since only the variance of $z$ matters in that respect, it is relevant to information-theoretic derivations and chiefly that of the spectral efficiency, which does depend on the distribution of $z$.

It is nevertheless customary to analyze $\Cexact$ in the form it would have if $\{ H_k \}_{k=1}^\infty$ were actually known by the receiver at the origin and $z$ were consequently Gaussian, namely the form
\begin{align}
\label{MWC16}
\!\!\Cub & = \mathbE_{ \{ \Hk \}_{k=0}^\infty} \!\! \left[\log_2\left(1+\frac{P \, \ro^{-\eta} |\Ho|^2 }{P \sum_{k=1}^{\infty} \rk^{-\eta} |\Hk|^2 + N_0} \right) \right]
\end{align}
where the tacit---and seldom made explicit---redefinition of $z$ as Gaussian is unmistakable from $I(\so;\sqrt{\gamma}\so+z) = \log_2(1+ \gamma)$, which holds only when $\so$ and $z$ are Gaussian. 
As it turns out, a Gaussian modeling of $z$ is not unreasonable because, 
if a decoder is intended for Gaussian interference-plus-noise (either by design or because the distribution thereof is unknown), then the spectral efficiency is precisely as if the interference-plus-noise were indeed Gaussian, regardless of its actual distribution \cite{lapidoth2002fading}.
And, once $z$ is taken to be Gaussian, the capacity-achieving signal distribution is also Gaussian, validating our choice for $\so$.
At the same time, the granting of $\{ \Hk \}_{k=1}^\infty$ as additional side information to the receiver renders (\ref{MWC16}) an upper bound to $\Cexact$, hence the denomination $\Cub$.

While much more tractable than $\Cexact$, the form of $\Cub$ has the issue of depending not only on $H_0$, but further on $\{ H_k \}_{k=1}^\infty$. This still clutters its analysis considerably, as will be seen later. 

Alternatively, what we propound in this paper is
to model $z$ as Gaussian, but forgoing the small-scale variations in its power, i.e., to use
\begin{align}
\label{pie2}
z \sim \mathcal{N}_\mathbb{C} \left( 0 , \, P \sum_{k=1}^{\infty} \rk^{-\eta} + N_0 \right) .
\end{align}
The closeness between this distribution for $z$ and its original brethren in (\ref{eq:aggr_z}) has been tightly bounded
\cite{1413204,ak2016gaussian}.

The model proposed in (\ref{pie2}) 
has the virtue of rendering $z$ Gaussian without the strain 
of gifting the receiver with $\{ \Hk \}_{k=1}^\infty$ and, as in other contexts where it has been tested \cite{Poisson-field-partI,MunMorLoz-TWC14,FrameworkD2D}, the result of this restrain turns out to be gratifyingly good for cellular network analysis.
With $z$ as in (\ref{pie2}), the instantaneous SINR then becomes
\begin{align}
\label{UoE}
\SINRo &= \frac{P \, \ro^{-\eta} \, |\Ho|^2 }{P \sum_{k=1}^{\infty} \rk^{-\eta} + N_0} ,
\end{align}
and the corresponding ergodic spectral efficiency is
\begin{align}
\label{H2020}
C = \mathbE_{\Ho} \! \left[\log_2\left(1+\frac{P \, \ro^{-\eta} |\Ho|^2 }{P \sum_{k=1}^{\infty} \rk^{-\eta} + N_0} \right) \right] ,
\end{align}
which is the quantity we shall work with. 



Since, with Gaussian codewords and a given variance for $z$, the mutual information is minimized when $z$ is Gaussian \cite{diggavi2001worst}, we have that $C \leq \Cexact$. 
The similarity between $C$ and $\Cexact$, with the former characterized analytically and the latter obtained through Monte-Carlo simulation, is illustrated throughout the paper. 

Contrasting the instantaneous SINR expressions in (\ref{eq:SIR_wrong}) and (\ref{UoE}), the analytical virtues of our model for $z$ become evident once we rewrite the latter as
$\SINRo = \rho \, |\Ho|^2$,
where
\begin{align}
\label{eq:localavgSIR}
\rho &= \frac{P \, \ro^{-\eta}}{P \sum_{k=1}^{\infty} \rk^{-\eta} + N_0}
\end{align}
is the local-average\footnote{The term ``local-average'' indicates averaging over the small-scale fading only. It follows that the ``local-average SINR'' is its average value over a region small enough for the path loss (and the shadowing, if applicable) not to change noticeably.}
SINR at the origin, fixed over any entire codeword and cleanly separated from the fluctuant term $|H_0|^2$; this reflects the decoupling between the large- and small-scale dependences. 
In interference-limited conditions ($P/N_0 \rightarrow \infty$), the local-average SINR specializes to $\rho = \ro^{-\eta} / \sum_{k=1}^{\infty} \rk^{-\eta} $.
For given BS and user locations, $\rho$ becomes determined and the conditional distribution of the instantaneous SINR is then given directly by that of $|\Ho|^2$, i.e., by the CDF $F_{|H_0|^2}(\cdot)$, while the ergodic spectral efficiency of a user with local-average SINR $\rho$ is
\begin{align}
C (\rho) &= \mathbE_{H_0} \Big[ \log_2 \big (1+ \rho \, |H_0|^2  \big) \Big] \\
\label{eq:SE_specific}
&= \int_{0}^{\infty} \log_2 (1 + \rho \, \xi) \,\mathrm{d} F_{|H_0|^2} (\xi) ,
\end{align}
which, through $\rho$, sets the stage for further computations involving the geometry of the network.
As anticipated, it is here where stochastic geometry can be applied with all its potency, undistracted by lingering small-scale terms.

\begin{exmp}
\label{Ex1}

In Rayleigh fading,
$F_{|H_0|^2}(\xi) = 1- e^{- \xi} $,
from which
\begin{align}
\label{eq:SE_spec_geo}
C(\rho) = e^{1/\rho} \, E_{1} \! \left( \frac{1}{\rho} \right) \log_2 e ,
\end{align}
where $E_n(x) = \int_{1}^{\infty} t^{-n} e^{-x \,t} \, {\rm d}t$ is an exponential integral.

\end{exmp}
For fading distributions other than Rayleigh, or with MIMO or other features, corresponding forms can be obtained for $C(\rho)$, always with the key property of these being a function of $\rho$ and not of the instantaneous fading coefficients.

\begin{exmp}

Let transmitter and receiver be equipped, respectively, with $\Nt$ and $\Nr$ antennas.
If $H_0$ is replaced by an $\Nr \times \Nt$ channel matrix $\boldsymbol{H}_0$ having IID Rayleigh-faded entries, then \cite{Shin-Lee-IT03}
\begin{align}
	\nonumber
\!\!\!C^{\scriptscriptstyle \mathsf{MIMO}}_{\scriptscriptstyle \Nr\times\Nt}(\rho) &= e^{\Nt/\rho} \sum_{i=0}^{m-1} \sum_{j=0}^{i} \sum_{\ell=0}^{2\,j} \Bigg\{ \binom{2\,i-2\,j}{i-j} \binom{2\,j+2\,n-2\,m}{2\,j-\ell} \\ 
&\qquad \qquad \qquad \cdot \frac{(-1)^{\ell} \, (2\,j)! \, (n-m+\ell)!}{2^{2i-\ell} \, j! \, \ell! \, (n-m+j)!} \sum_{q=0}^{n-m+\ell} E_{q+1} \left(\frac{\Nt}{\rho}\right) \Bigg\} \, \log_2 e
	\label{halloween}
\end{align}
with $m = \min(\Nt,\Nr)$ and $n = \max(\Nt,\Nr)$.

\end{exmp}

\begin{exmp}
\begin{figure}
	\centering
	\includegraphics [width=0.75\columnwidth]{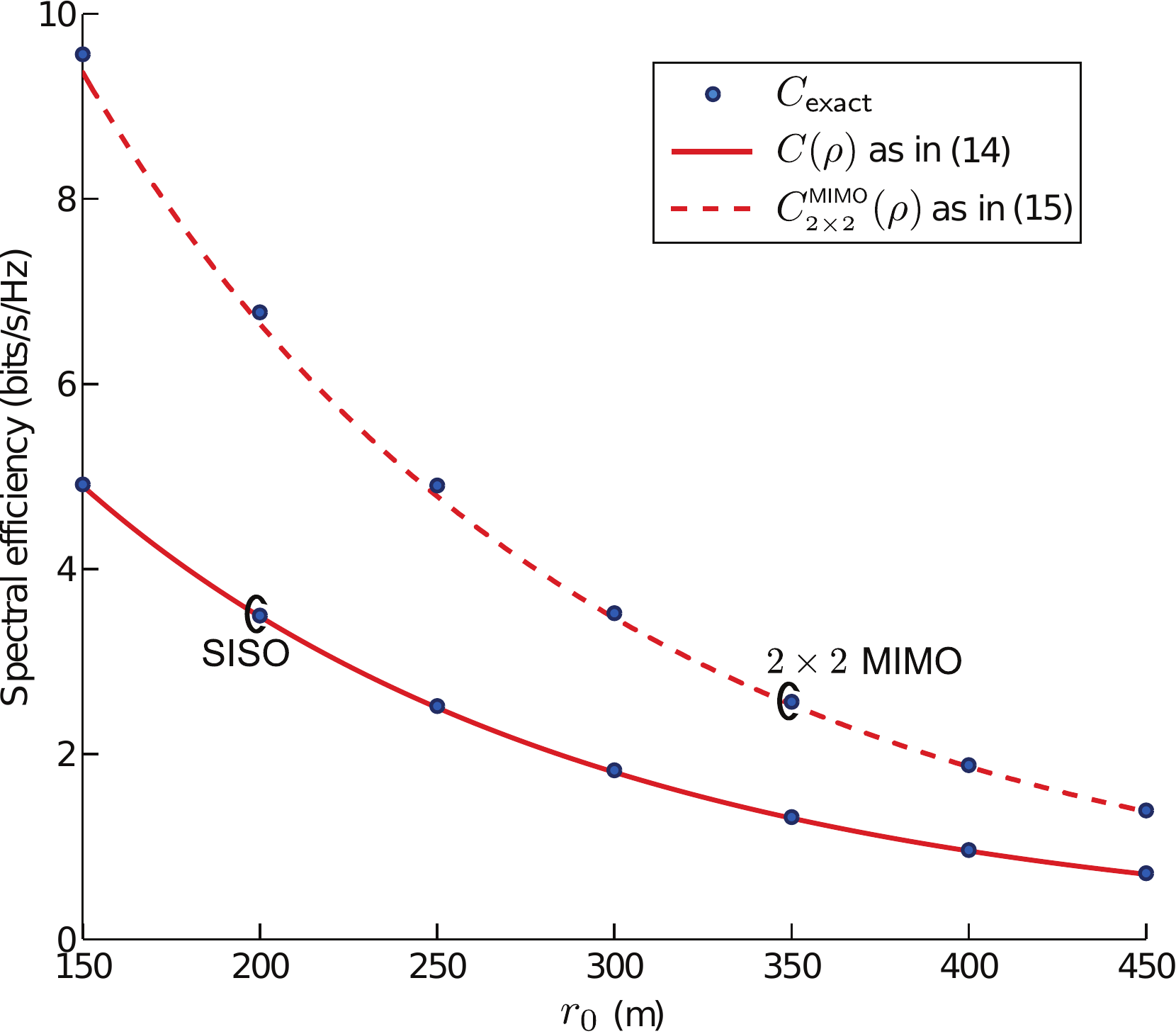} 
	\caption{Spectral efficiency vs.~$\ro$ for $\lambdab=2\, \textnormal{BSs}/{\textnormal{km}}^2$, $\eta=3.8$
		and $\rk = \Gamma(k+1.5)/(\sqrt{\pi \lambdab}\, \Gamma(k+1))$ for $k=1, \ldots, 100$. The $99\%$ confidence intervals around $\Cexact$
		range from $\pm 0.029$ at $\ro = 150$ m to $\pm 0.009$ at $\ro = 450$ m, for both SISO and MIMO.}
	\label{Fig:fig0}
\end{figure}	
Let us consider the application of (\ref{eq:SE_spec_geo}) to an interference-limited network with 100 interfering BSs. To typify the network, we set $\rk$ to the expected value of the distance to the $k$th nearest point in a PPP: 
\[\rk =\frac{\Gamma(k+1.5)}{\sqrt{\pi \lambdab}\, \Gamma(k+1)}\] 
for $k=1, \ldots, 100$~\cite{Haenggi05}. We further set $\eta=3.8$ and $\lambdab=2$ BSs/km$^2$ and neglect the noise. Shown in Fig. \ref{Fig:fig0} is $C(\rho)$ in (\ref{eq:SE_spec_geo})
compared against $\Cexact$.
The same comparison is provided for MIMO with $\Nt=\Nr=2$.
In both cases the differences are minute, supporting the interference modeling approach propounded in this paper.
\end{exmp}

\section{Distribution of the Local-Average SINR}

Thus far, the link distances $ \{ \rk \} $ have been conditioned upon. Once they are released, $\rho$ becomes itself a random variable whose distribution is induced by the process $\Phib$. The corresponding CDF, $F_{\rho}(\cdot)$, is to be a central ingredient in our analysis of the ergodic spectral efficiency. 
This important function, extensively utilized by system designers, has traditionally been obtained by means of simulation over lattice networks with shadowing and random translations \cite{zorzi1997analytical}.
For hexagonal cells in particular, and without shadow fading, an infinite series solution is also available~\cite{hexagonal-analytical}.

Here we set out to characterize $F_\rho(\cdot)$ for Poisson networks, so as to implicitly incorporate shadowing, and specifically for interference-limited Poisson networks.
Interestingly, with PPP-populated BSs, the powers received by any given user are statistically invariant---save for a scaling of the BS density if noise were not negligible---to the distribution of the channel gains  \cite{4675736,ShadowingPPP,5678764} and thus
$F_\rho(\cdot)$ is equivalent to what $F_{\SIRo}(\cdot)$ would look like if users connected to the BS with the strongest instantaneous link. 
This distribution was established
in \cite{Dhillon-HetNet-JSAC12,blaszczyszyn2013using,6120366,7305791}, very compactly for arguments above $1$ and in a still manageable form for arguments between $1/2$ and $1$, but in an accelerating cumbersome fashion (involving progressively higher-dimensional integrations)
as the argument of the CDF dips below $1/2$ \cite[Sec. V.A]{6120366}, \cite[Cor. 19]{7305791}.
As alternatives, \cite{7243344,blaszczyszyn2013using,6933943} derived $F_{1/\rho}(\cdot)$ in the Laplace domain, which would then require numerical inversion,
while \cite{ICC16-Ganti-Haenggi} showed that the lower tail ($\theta \rightarrow 0$) of
$F_\rho(\theta)$ satisfies
\begin{align}
\label{eq:SIR_asymptot}
\log F_\rho(\theta) = \frac{s^\star}{\theta} + o(1)
\end{align}
with $s^\star<0$ being the solution to 
\begin{align}
\label{saul}
s^{\star\delta}\,\bar\Gamma(-\delta,s^\star)=0
\end{align}
where $\bar\Gamma$ is the lower incomplete gamma function
and, for compactness, we have introduced the shorthand notation $\delta = 2/\eta$. Since it only depends---through $\delta$---on the path loss exponent $\eta$, the parameter $s^\star$ can be precomputed for all relevant values thereof by solving (\ref{saul}) using any standard software package. 
Values of $s^\star$ for some typical $\eta$ are listed in Table \ref{tab:sstarvalues}.

\begin{table}
	\caption{Parameter $s^\star$ and the corresponding $\delta$ for typical values of the path loss exponent $\eta$.}
	\label{tab:sstarvalues}
	\vspace{-0.1in}
	\begin{center}
		\begin{tabular}{ |c|c|c|}
			\hline
			$\eta$ & $\delta$ & $s^\star$  \\ \hline\hline
			3.5 & 0.571 & -0.672 \\ \hline
			3.6 & 0.556 & -0.71 \\ \hline
			3.7 & 0.540 & -0.747 \\ \hline
			3.8 & 0.526 & -0.783\\ \hline
		\end{tabular}
		\begin{tabular}{ |c|c|c|}
			\hline
			$\eta$ & $\delta$ & $s^\star$  \\ \hline\hline
			3.9 & 0.513  & -0.819\\ \hline
			4  & 0.5 & -0.854\\ \hline
			4.1 & 0.488 & -0.888\\ \hline
			4.2 & 0.476 & -0.922 \\ \hline
		\end{tabular}
	\end{center} 
	\vspace{-0.4cm}
\end{table}

The approach we take is to apply the exact form 
down to $\theta = 1/2$ and then (\ref{eq:SIR_asymptot}) for $\theta < 1/2$.
This combination gives
\begin{align}
\label{eq:SIRCDF_asymptotic1}
			 \left\{ \begin{array}{l l}
			\!\! F_{\rho}(\theta) \simeq e^{s^\star/\theta} \qquad &  0 \leq \theta < \frac{s^\star}{\log A_\delta} \\
			\!\! F_{\rho}(\theta) \approx A_\delta  \qquad &   \frac{s^\star}{\log A_\delta} \leq \theta < 1/2 \\
			\!\! F_{\rho}(\theta) = 1 - \theta^{-\delta} \sinc \, \delta + B_{\delta} \! \left(\frac{\theta}{1-\theta}\right)  \qquad &   1/2 \leq \theta < 1 \\
			\!\!F_{\rho}(\theta) = 1 - \theta^{-\delta} \sinc \, \delta \qquad &   \theta \geq 1,
						\end{array} \right.
\end{align}
where "$\simeq$" indicates an approximation with asymptotic ($\theta \to 0$) equality while
\begin{align}
A_\delta = 1 - 2^{\delta} \sinc \, \delta +  B_{\delta} (1)
\end{align}
and
\begin{align}
B_{\delta}(x) &= \frac{\delta \, \sinc^2(\delta)\, {\Gamma^2 (\delta + 1 )} \, {}_2 F_1 \big(1, \delta+1; 2 \, \delta + 2; -1/x \big) }{x^{1+2\,\delta} \; \Gamma (2\,\delta + 2 )} 
\end{align}
with ${}_2 F_1$ the Gauss hypergeometric function.
When the path loss exponent is $\eta = 4$, we have that $\delta=1/2$ and the above specialize to
\begin{align}
\label{merdaCUP}
\left\{
\begin{array}{l l}
			\!\!F_{\rho}(\theta) \simeq e^{-0.854/\theta} \qquad\quad &  0 \leq \theta < 0.457 \\
			\!\!F_{\rho}(\theta) \approx 0.154 \qquad\quad &    0.457 \leq \theta < 1/2 \\
			\!\! F_{\rho}(\theta) = 1 - \frac{4 \sqrt{\theta} - \theta - 1}{\pi \, \theta}  \qquad\quad &   1/2 \leq \theta < 1 \\
			\!\! F_{\rho}(\theta) = 1 - \frac{2}{\pi \sqrt{\theta}} \qquad\quad &   \theta \geq 1 .
			\end{array}
\right.
\end{align}

An even simpler, slightly less accurate expression for $F_\rho(\cdot)$ is obtained using the exact form only down to $\theta=1$ while stretching the lower tail expansion in (\ref{eq:SIR_asymptot}) up to $\theta=1$. This gives
\begin{align}
\label{eq:SIRCDF_asymptotic2}
 \left\{ \begin{array}{l l}
				\!\! F_{\rho}(\theta) \simeq e^{s^\star/\theta} \qquad\quad &  0 \leq \theta < \frac{s^\star}{\log (1 -\sinc \, \delta)} \\
				\!\! F_{\rho}(\theta) \approx 1 -\sinc \, \delta  \qquad\quad &   \frac{s^\star}{\log (1 -\sinc \, \delta)} \leq \theta < 1 \\
				\!\! F_{\rho}(\theta) = 1 - \theta^{-\delta} \sinc \, \delta \qquad\quad &   \theta \geq 1 .
						\end{array} \right.
\end{align}

%
%

\section{Distribution of the Spectral Efficiency}
\label{SectionSE1}


The randomness that $\rho$ acquires once the BS positions are randomized is then inherited by $C$, and the corresponding CDF
provides a complete description of the ergodic spectral efficiency offered by the network over all locations \cite{Poisson-field-partII}.
By mapping the applicable function $C(\rho)$ onto the expressions for $F_\rho(\cdot)$ put forth in the previous section, $F_C(\cdot)$ is readily characterized. In interference-limited conditions, such $F_C(\cdot)$ depends only on the path loss exponent, $\eta$.

\subsection{SISO}

In Rayleigh-faded SISO channels, $C(\rho)$ is given by (\ref{eq:SE_spec_geo}). By resorting to the invertible approximation~\cite{Catreux-erg-seff}
\begin{align} \label{eq:Erg Seff approx}
e^{ \nu } E_1 (\nu) \log_2 e \approx  1.4 \log \left( 1+ \frac{0.82}{\nu} \right) 
\end{align}
it becomes possible to write $\rho \approx \frac{e^{C/1.4}-1}{0.82}$ and subsequently, by means of (\ref{eq:SIRCDF_asymptotic1}),
\begin{align}
\label{eq:Erg Seff CDF approx}
\!\! F_{C} (\gamma) &\approx F_{\rho} \left(  \frac{e^{\frac{\gamma}{1.4}} -1 }{0.82} \right)
\end{align}
which can be expressed 
\begin{align}
\label{eq:CDF_SE_asym}
\!\! F_{C} (\gamma) \approx 
\left\{ \begin{array}{l l}
 e^{\frac{0.82 \, s^\star}{\exp(\gamma/1.4)-1}} \qquad\quad &  0 \leq \gamma < 1.4 \log \! \left(1+ \frac{0.82 \, s^\star}{\log \! A_\delta }\right) \\
 A_\delta  \qquad\quad &   1.4 \log \! \left(1+ \frac{0.82 \, s^\star}{\log \! A_\delta }\right) \leq \gamma <  0.48 \\
 1 -  \frac{\sinc \, \delta}{\left(  \frac{e^{\gamma/1.4} -1 }{0.82}  \right)^{\delta}} + B_{\delta}\left(\frac{e^{\gamma/1.4} -1 }{1.82 - e^{\gamma/1.4}} \right)  \qquad\quad &   0.48 \leq \gamma < 0.84 \\
 1 -  \frac{\sinc \, \delta}{\left(  \frac{e^{\gamma/1.4} -1 }{0.82}  \right)^{\delta}} \qquad\quad &   \gamma \geq 0.84,
 \end{array} \right.
\end{align}
whose accuracy is validated in the following example.

\begin{figure}
	\centering
	\includegraphics [width=0.75\columnwidth]{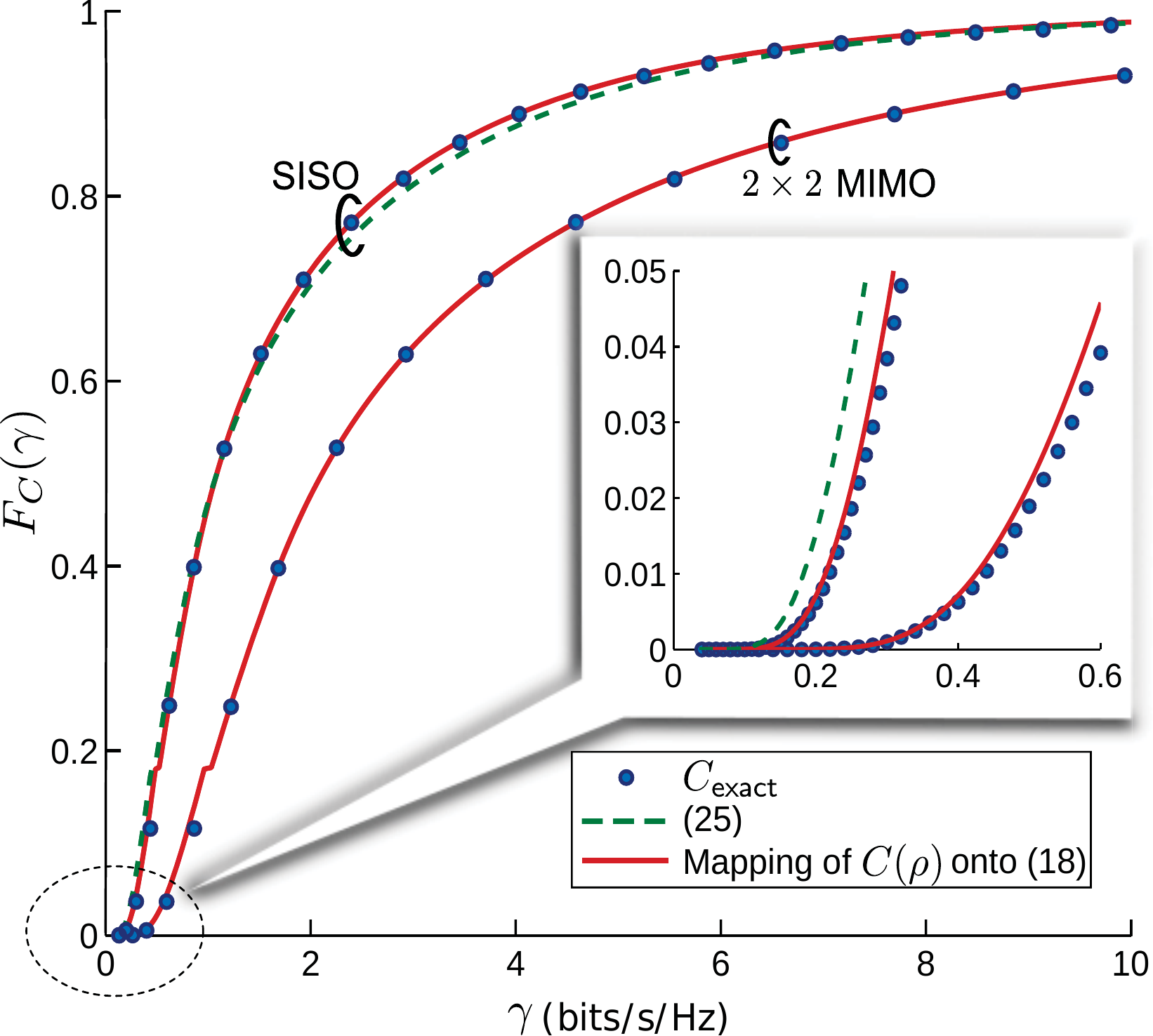}
	\caption{CDF of ergodic spectral efficiency for $\eta=3.8$. In the inset, a zoom-in of the lower tail.}
	\label{Fig:val2}
\end{figure}
\begin{exmp}
\label{SECDF_valid}
Consider an interference-limited network with $\eta=3.8$.
Shown in Fig. \ref{Fig:val2} are (\ref{eq:CDF_SE_asym}), as well as the numerical mapping of $C(\rho)$---without the bypass of its invertible approximation---onto (\ref{eq:SIRCDF_asymptotic1}), both solutions contrasted against $\Cexact$. For the computation of $\Cexact$ via Monte-Carlo, here and in all examples involving simulations hereafter, PPP-populated BSs (1000 on average) are dropped in a disk centered at the receiver.
\end{exmp}

For $\eta=4$, (\ref{eq:CDF_SE_asym}) specializes to
\begin{align}
\label{esterilla}
F_{C} (\gamma)\! \approx \! \left\{ \begin{array}{l l}
			\!\!\! e^{- \frac{0.7}{\exp(\gamma/1.4)-1}} \quad &  0 \leq \gamma < 0.44 \\
			\!\!\! 0.154   \quad &    0.44 \leq \gamma <  0.48 \\
			\!\!\! 1 - \frac{4}{\pi} \sqrt{\frac{0.82}{e^{\gamma/1.4}-1}} + \frac{e^{\gamma/1.4}-0.18}{\pi \, (e^{\gamma/1.4}-1)}  \quad &  0.48 \leq \gamma < 0.84 \\
			\!\!\! 1 - \frac{2}{\pi} \sqrt{\frac{0.82}{e^{\gamma/1.4}-1}}  \quad &   \gamma \geq 0.84 .
			\end{array} \right.
\end{align}

In contrast to these pleasing results, without the model for $z$ propounded in this paper the distribution of the spectral efficiency over the network locations is far more inaccessible. Indeed, the corresponding $\Cub$ can be rewritten (cf. App. \ref{clasico}) as
\begin{align}
\label{eq:SE_wrong}
\Cub &= \log_2 e \int_{0}^{\infty} \frac{e^{-x \, \ro^\eta \, N_0/P}}{1+x} \prod_{k=1}^{\infty} \frac{1}{1+ x \left(\ro/\rk\right)^\eta} \, {\rm d}x,
\end{align}
which is no longer a function of singly $\rho$, whose distribution was established in the previous section;
rather, (\ref{eq:SE_wrong}) is a more involved function of $\{ \rk \}_{k=0}^\infty$ and offers no obvious way of disentangling these dependences. Faced with this obstacle,
some authors choose
to instead characterize the distribution of
\begin{align}\nonumber
\log_2 \!\left( 1+\frac{P \, \ro^{-\eta} |\Ho|^2 }{P \sum_{k=1}^{\infty} \rk^{-\eta} |\Hk|^2 + N_0} \right)
\end{align}
over $\{ H_k \}_{k=0}^\infty$ as well as $\{ \rk \}_{k=0}^\infty$ \cite{DL-MIMO-LB}\cite[Sec. VII.A]{stochgeo-tutorial-ElSawy}. 
However, the mixing of small- and large-scale variations within this quantity clutters potential observations. Moreover, the generalization to more involved settings, say MIMO, appears arduous or outright hopeless. 
Indeed, existing stochastic geometry analyses of spectral efficiency featuring MIMO are restricted to beamforming or space-division multiple access, rather than spatial multiplexing \cite{dhillon2013downlink,di2015stochastic,MIMO-TRXDiv-stochgeo,MIMO-reTX-stochgeo,MUMIMO-Li-TCOM16,DL-MIMO-LB}.

\subsection{MIMO}

Our approach, in contrast, only requires mapping the appropriate $C(\rho)$ onto  $F_\rho(\cdot)$. Whenever $C(\rho)$ does not lend itself to inversion, even approximately, it is straightforward to perform this mapping numerically.

\begin{exmp}
	\label{SECDF_valid2}
Reconsider Example \ref{SECDF_valid}, but now with $\Nt=\Nr=2$ such that, from (\ref{halloween}),
\begin{align}
\Cmimo(\rho) & = 2 \, e^{2/\rho} \left[E_1 \! \left(\frac{2}{\rho}\right)+E_3 \! \left(\frac{2}{\rho}\right)\right] \log_2 e \label{eq:MIMO SE form1} \\
 &= \left[2 \, e^{2/\rho} \, E_1 \! \left(\frac{2}{\rho}\right) \, \left(1+\frac{2}{\rho^2}\right) +  \left(1-\frac{2}{\rho}\right) \right] \log_2 e.
\label{eq:MIMO SE}
\end{align}
Fig.~\ref{Fig:val2} depicts the numerical mapping of $\Cmimo(\rho)$ onto (\ref{eq:SIRCDF_asymptotic1}), as well as the corresponding $\Cexact$.
\end{exmp}

\subsection{Lognormal Fit}

Inspecting the distribution of $C(\rho)$ we observe that, interestingly, it closely resembles a lognormal function, i.e., that
$\log C(\rho)$ admits a rather precise Gaussian fit. This opens the door to an alternative to $F_C(\cdot)$ as obtained by mapping $C(\rho)$ onto $F_\rho(\cdot)$,
namely the alternative
$\log C(\rho) \sim \mathcal{N}(\mu,\sigma^2)$ with
\begin{align}
\label{AG1}
\mu & = \int_0^\infty \log C(\theta) \, \mathrm{d}F_\rho(\theta) \\
\sigma^2 & = \int_0^\infty \big[ \log C(\theta) \big]^2 \, \mathrm{d}F_\rho(\theta) - \mu^2 .
\label{AG2}
\end{align}

\begin{exmp}

Let $\eta=4$.
For $\Cmimo(\rho)$ as given in (\ref{eq:MIMO SE}) and $F_\rho(\cdot)$ as given in (\ref{merdaCUP}),
the numerical integrations in (\ref{AG1})--(\ref{AG2}) yield $\mu=0.92$ and $\sigma^2=0.8$.
Fig.~\ref{fig:logn} presents the empirical PDF of $\log \Cmimo(\rho)$, generated via Monte-Carlo, and a Gaussian PDF with $\mu=0.92$ and $\sigma^2=0.8$.

\begin{figure} 
\centering
\includegraphics [width=0.75\columnwidth]{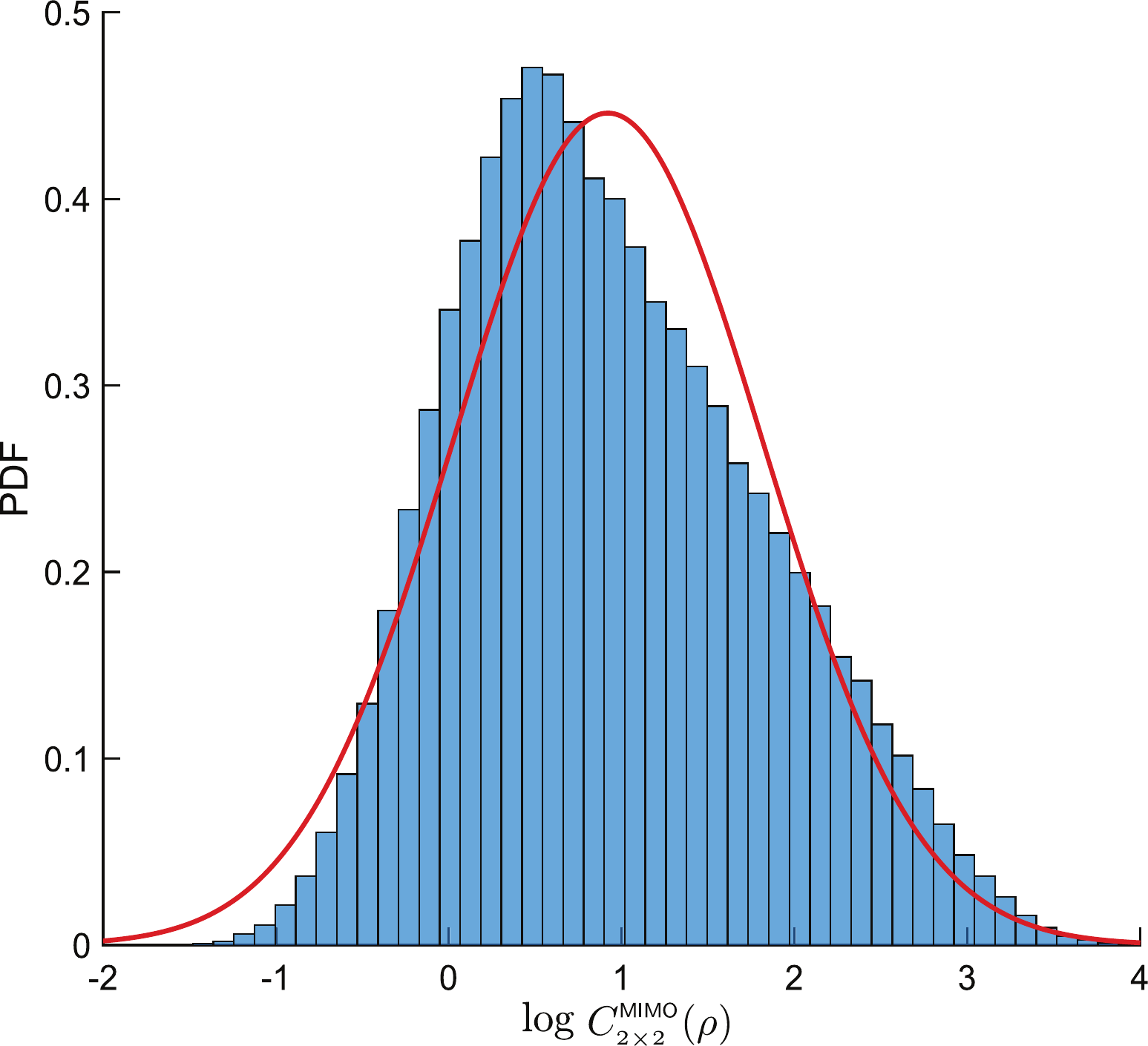}
\caption{Empirical PDF of the logarithm of the $2 \times 2$ MIMO ergodic spectral efficiency for $\eta=4$ and its normal fit.
}
\label{fig:logn}
\end{figure}

\end{exmp}

Thanks to this lognormal behavior, which holds for both SISO and MIMO, a Gaussian distribution with mean $\mu$ and variance $\sigma^2$ can provide
a quick idea of the disparity of the user experiences throughout the network.
This complements the coverage analysis presented in the next section, which is more precise but valid only for the lower tail. The lognormal fit, rather,
may be used to determine the fraction of users whose spectral efficiency lies within a certain interval of the average, or it may simplify further calculations that require averaging with respect to the distribution of $C(\rho)$.

\section{Coverage}

While the entire CDF of ergodic spectral efficiency is relevant from a network-design perspective, the lower tail is especially important as it determines the coverage, i.e., the share of network locations in which a minimum level of service can be provided. The shape of the lower tail, in particular, reveals the improvement in coverage with a diminishing service requirement or, equivalently, the sacrifice in coverage that is necessary to guarantee a better minimum service.
The behavior of this tail (which, for $\eta=3.8$, is detailed in the inset of Fig. \ref{Fig:val2}) warrants attention, and this section is devoted precisely to its analysis and to gleaning coverage insights from it.

\subsection{SISO}

For SISO, the CDF lower tail is seen from (\ref{eq:Erg Seff CDF approx}) to behave as
\begin{equation}
F_{C} (\gamma) \approx e^{\frac{0.82 \, s^\star}{\exp(\gamma/1.4)-1}} .
\end{equation}
Using
$
e^{\gamma/1.4}-1 = 0.72 \, \gamma + \mathcal{O}(\gamma^2),
$
the tail behavior can be re-expressed as
\begin{equation}
\label{MuchRain}
F_{C} (\gamma) \approx e^{1.15 \, s^\star/\gamma} 
\end{equation}
where, recall, $s^\star$ is negative.
By inverting (\ref{MuchRain}), the spectral efficiency $\gamma$ achievable on a share $1- \xi$ of the network is seen to satisfy (for $\xi \leq A_\delta$)
\begin{equation}
\label{iMuchRain}
\gamma \approx \frac{1.15 \, s^\star}{\log_e \xi} .
\end{equation}

For $\eta=4$, we can recall $s^\star = -0.854$ and the above direct and inverse expressions then specialize, respectively, to
\begin{equation}
\label{Topillo}
F_{C} (\gamma) \approx e^{-1/\gamma} 
\end{equation}
and (for $\xi \leq 0.154$)
\begin{equation}
\gamma \approx \frac{1}{\log_e 1 / \xi} .
\end{equation}

\begin{exmp}
For SISO and $\eta=4$, (\ref{Topillo}) is depicted in Fig. \ref{UPFUPC} alongside previously obtained expressions for $F_C(\cdot)$ as well as the exact tail computed via Monte-Carlo.
\end{exmp}

\begin{exmp}
\label{GoPro}
For SISO and $\eta=4$, the spectral efficiency $\gamma$ achievable in $99\%$ of the network satisfies
\begin{align}
\gamma \approx \frac{1}{\log_e 100} = 0.22
\end{align}
whereas the exact value, obtained numerically, equals $0.24$ bits/s/Hz.
\end{exmp}

\subsection{MIMO}

With $\Nt$ transmit and $\Nr$ receive antennas, the ergodic capacity expands as \cite{lozano2003multiple}
\begin{equation}
C(\rho) = \Nr \, \rho \log_2 e + \mathcal{O}(\rho^2) 
\end{equation}
indicating a linear scaling with $\Nr$ and no dependence on $\Nt$. This low-SNR behavior is very robust, holding irrespectively of the fading distribution and in the face of antenna correlation, signifying that what matters in this regime is only the receiver's ability to capture power.

Applying the linear scaling of the low-SNR capacity with $\Nr$ to (\ref{MuchRain}) and (\ref{iMuchRain}) we obtain
\begin{equation}
F_{C} (\gamma) \approx e^{1.15 \, s^\star \Nr/\gamma} 
\label{Topillo2}
\end{equation}
and
\begin{equation}
\gamma \approx \frac{1.15 \, s^\star \Nr}{\log_e \xi} .
\end{equation}
which can be readily specialized to $\eta=4$ as well.

\begin{figure} 
\centering
\includegraphics [width=0.75\columnwidth]{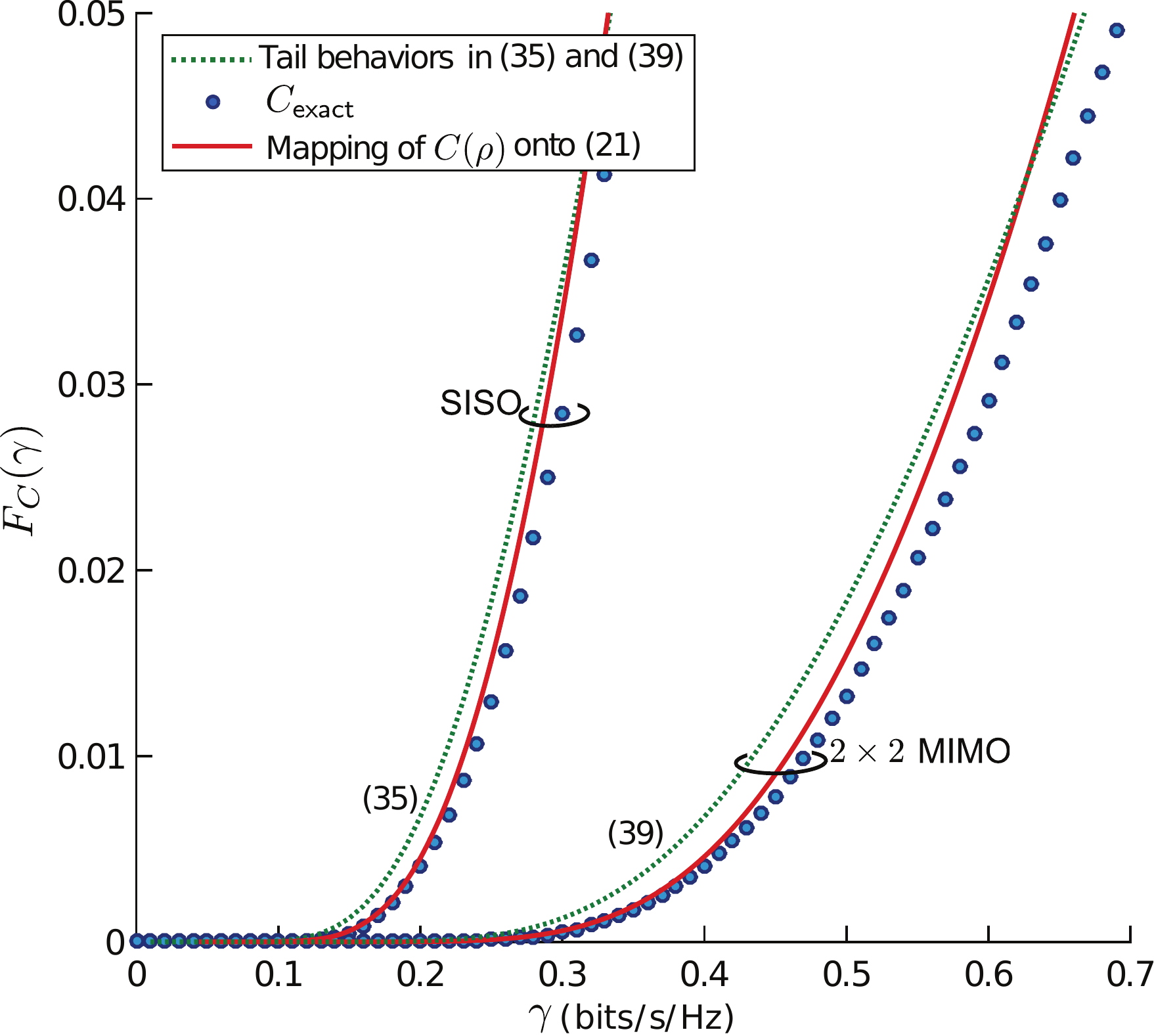}
\caption{Lower tail of the ergodic spectral efficiency CDF for $\eta=4$.}
\label{UPFUPC}
\end{figure}

\begin{exmp}
For $\Nt=\Nr=2$ and $\eta=4$, (\ref{Topillo2}) is depicted in Fig. \ref{UPFUPC} alongside previously obtained expressions for $F_C(\cdot)$ as well as the exact tail computed via Monte-Carlo.
\end{exmp}

Since coverage is not gained or lost in reference to small-scale fades, which are highly localized in space, time and frequency,
care must be exercised not to infer coverage from the distribution of instantaneous SINR or of instantaneous spectral efficiency, both of which are dominated by small-scale fading. With Rayleigh fading and $\eta=4$ in particular, the distribution of the instantaneous SIR throughout the network equals \cite{Andrews-Baccelli-Ganti-coverage}
\begin{equation}
F_{\sf SIR}(\theta) = 1 - \frac{1}{1+\sqrt{\theta} \arctan \big( \sqrt{\theta} \big)}
\end{equation}
whose lower tail behaves as $F_{\sf SIR}(\theta) = \theta + \mathcal{O}(\theta^2)$. This linear decay is drastically different from the exponential ones derived in this section on the basis of $F_C(\cdot)$. At $99\%$ coverage, $F_{\sf SIR}(\cdot)$ would map to a value of $0.014$ bits/s/Hz, far from the exact value of $0.24$ and from the value of $0.22$ we obtained in Example \ref{GoPro}.
It is therefore important that coverage be gleaned from the ergodic spectral efficiency, which is impervious to small-scale fluctuations, rather than from quantities subject to those fluctuations.

\section{Spatial Average of the Spectral Efficiency}
\label{SectionSE2}

Sometimes, it is of interest to condense $F_C(\cdot)$ down to a single quantity, and in that case the average
is the logical choice.
Under spatial ergodicity, which holds for the PPP and many other point processes \cite{MHaenggi12}, this quantity equals the average of all per-user spectral efficiencies in any realization of the network.

Here again, the approach propounded in this paper proves advantageous.
For any setting for which $C(\rho)$ is available, our expressions for $F_{\rho} (\cdot)$ enable computing
\begin{align}\label{eq:SE avg gen}
\Cbar &= \int_{0}^{\infty} C(\theta) \, {\rm d} F_{\rho} (\theta),
\end{align}
which, in interference-limited conditions, again depends only on the path loss exponent.
Remarkably, the above integration can be solved for SISO and for (at least some) MIMO settings, and the expressions obtained involve only readily computable special functions. Although devoid of insight, these expressions allow circumventing large-scale Monte-Carlo simulations. 

\subsection{SISO}

In Rayleigh-faded SISO channels, with (\ref{eq:SE_spec_geo}) and (\ref{eq:SIRCDF_asymptotic2}) plugged into (\ref{eq:SE avg gen}), the integration yields (cf. App. \ref{siso_avg})
\begin{align}
 \Cbar  &\approx \frac{-s^\star \, \log_2 e}{1+s^\star} \, \left[ E_1\left(- \frac{s^\star}{D_\delta}\right) - e^{\frac{1+s^\star}{D_\delta}} \, E_1\left(\frac{1}{D_\delta}\right) \right] + \frac{\sin (\pi\delta) \, \log_2 e}{\pi} \, G^{2,2}_{2,3} \! \left(1 \Biggl | \Biggr. \begin{array}{l l} 0,1-\delta \\ 0,0,-\delta \end{array} \right) \label{eq:Avg seff asymp}
\end{align}
where
\begin{align}
G^{m,n}_{p,q}\left(z \Biggl | \Biggr. \begin{array}{l l} a_1,...,a_n,a_{n+1},...,a_p \\ b_1,...,b_m,b_{m+1},...,b_q \end{array} \right)
\end{align}
is the Meijer-G function while $D_\delta = s^\star/\log (1-\sinc \, \delta)$. An even more precise, albeit also more involved expression for $\Cbar$ can be obtained using (\ref{eq:SIRCDF_asymptotic1}) in lieu of (\ref{eq:SIRCDF_asymptotic2}).


\begin{exmp}
\label{SECDF_valid3}

Fig. \ref{Fig:val3} compares (\ref{eq:Avg seff asymp}) against $\barCexact$, the Monte-Carlo average of $\Cexact$, for $0.48 \leq \delta \leq 0.57$ corresponding to
 $3.5 \leq \eta \leq 4.2$.  For $\eta=4$ in particular,
\begin{align}
\Cbar & \approx 0.187 + \frac{\log_2 e}{\pi} \, G^{2,2}_{2,3} \left(1 \Biggl | \Biggr. \begin{array}{l l} 0,1/2 \\ 0,0,-1/2 \end{array} \right) 
= 1.99,
\end{align}
while $\barCexact$ is $2.01$
 
\begin{figure} 
\centering
 \includegraphics [width=0.75\columnwidth]{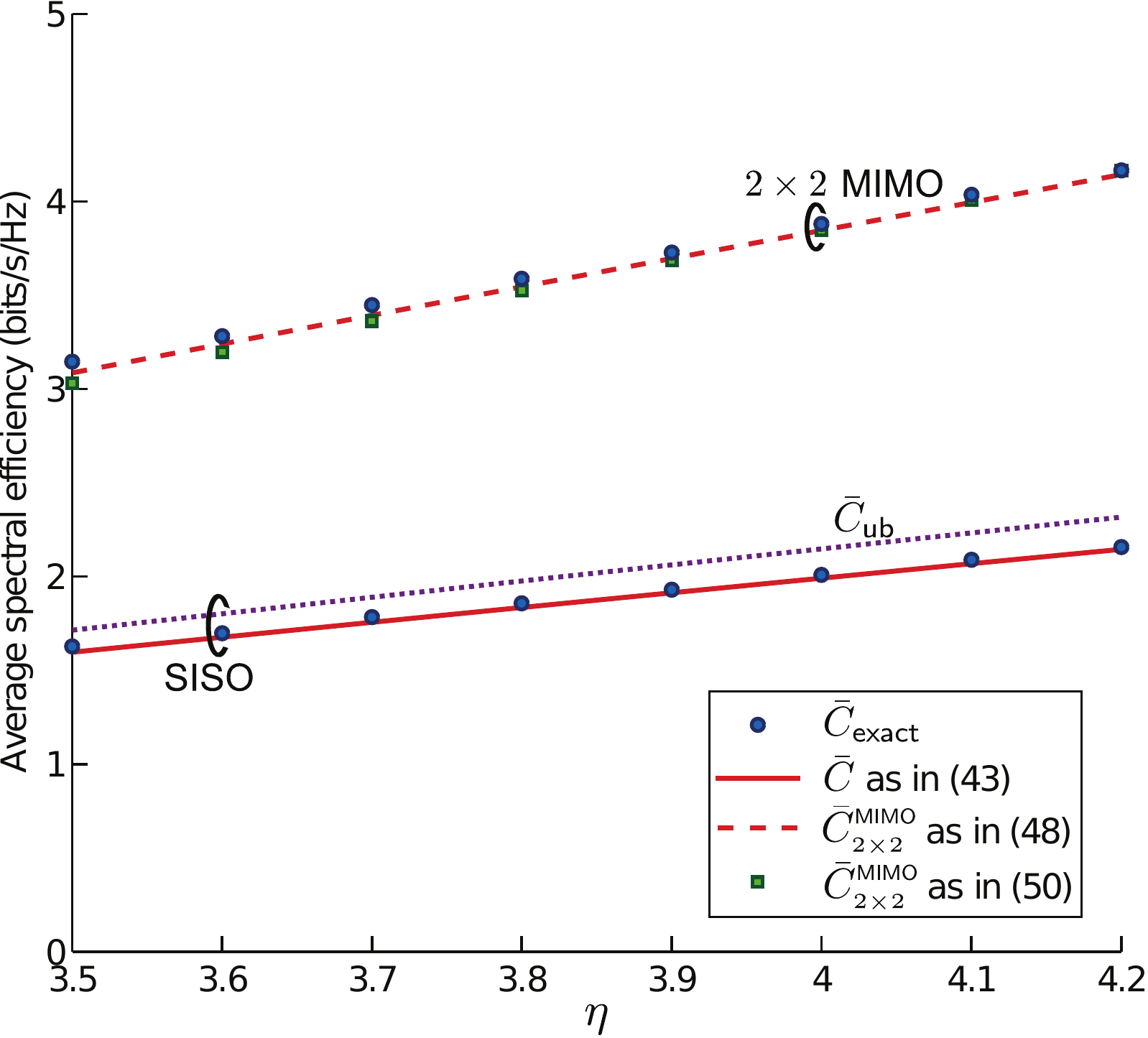}
\caption{Spatially averaged ergodic spectral efficiency as a function of $\eta$ for SISO and for $2 \times 2$ MIMO. The $99\%$ confidence interval around the Monte-Carlo results ranges from $\pm 0.008$ at $\eta = 3.5$ to $\pm 0.01$ at $\eta = 4.2$ for SISO and from $\pm 0.015$ at $\eta = 3.5$ to $\pm 0.02$ at $\eta = 4.2$ for MIMO. The SISO upper bound $\barCub$ is computed via (\ref{eq:Avg seff user no noise4}).}
\label{Fig:val3}
\end{figure}

\end{exmp}


It is worthwhile to contrast (\ref{eq:Avg seff asymp}) with its counterpart obtained without the model for $z$ propounded in this paper, namely the average of $\Cub$ given by \cite{Andrews-Baccelli-Ganti-coverage}
\begin{align}\label{eq:Avg seff user no noise3}
\barCub &= \int_{0}^{\infty} \frac{\log_2 e}{1+ (e^t-1)^{\delta} \int_{(e^t-1)^{-\delta}}^{\infty} \frac{1}{1+x^{1/\delta}} {\rm d}x} {\rm d}t
\end{align}
or, by means of Pfaff's transformation \cite{koepfhypergeometric}, equivalently by
\begin{align}\label{eq:Avg seff user no noise4}
\barCub &= \int_{0}^{\infty} \frac{\log_2 e}{{}_2 F_1 \! \left( 1, 1; 1 - \delta; \frac{\gamma}{1+\gamma}\right)} \, {\rm d}\gamma .
\end{align}

\begin{exmp}
Included in Fig. \ref{Fig:val3}, alongside its SISO counterparts $\Cbar$ and $\barCexact$, is also $\barCub$.
\end{exmp}

Besides being further from $\Cexact$ than our solution $\Cbar$, and requiring either a double integration or a single integral over a hypergeometric function, neither of the expressions for $\barCub$ offers a viable path to MIMO generalization.
With our approach, in contrast, the analysis of the spatial average becomes feasible also with MIMO.

\subsection{MIMO}

For $2\times2$ MIMO, the integration of $\Cmimo(\rho)$ as given in (\ref{eq:MIMO SE}) over $F_\rho(\cdot)$ as given in (\ref{eq:SIRCDF_asymptotic2}) returns (cf. App. \ref{siso_avg})
\begin{align}
\label{eq:MIMO avg SE1}
\Cbar^{\scriptscriptstyle \mathsf{MIMO}}_{\scriptscriptstyle 2\times2} &\approx \frac{2 s^\star e^{\frac{2+s^\star}{D_\delta}} \big[2 (2+s^\star)^2 - 4 (2+s^*) D_\delta + \big( 8+s^\star (4+s^\star) \big) D^2_\delta \big] \, E_1(2/D_\delta) }{(2+s^\star)^3 D^2_\delta \, \log 2} \spazio  \\
\nonumber
& \quad + \frac{s^\star D_\delta \big[2 \big(8+s^\star (4+s^\star) \big) D_\delta \, E_1 (-s^\star/D_\delta) - e^{s^\star/D_\delta} (2+s^\star) \big( (6+s^\star) D_\delta - 2(2+s^\star) \big)  \big] }{(2+s^\star)^3 D^2_\delta \, \log 2} \\
& \quad +\frac{\sin (\pi\delta) \log_2 e}{\pi} \, \left[ 2 \, G^{2,2}_{2,3} \left(1 \Biggl | \Biggr. \begin{array}{l l} 0,1-\delta \\ 0,0,-\delta \end{array} \right) + 4 \,  G^{2,2}_{2,3} \left(1 \Biggl | \Biggr. \begin{array}{l l} 0,-1-\delta \\ 0,0,-2-\delta \end{array} \right) + \frac{1 - \delta}{(1 + \delta) \, \delta} \right] . \nonumber  
\end{align}

An alternative expression not involving the Meijer-G function can be obtained using 
\[E_3(x) \approx e^{-3 x/2}/2\] 
and 
\[2 \, e^{2/\rho} E_1 \! \left(2/\rho\right) \log_2 e \approx 2.8 \, \log ( 1+ 0.41 \, \rho )\] 
to simplify $\Cmimo(\rho)$ in~(\ref{eq:MIMO SE form1}) into
\begin{align}\label{eq:MIMOSE_approx}
\Cmimo(\rho) \approx 2.8 \, \log ( 1+ 0.41 \, \rho ) + e^{-1 / \rho} \log_2 e.
\end{align}
With this form in place of (\ref{eq:MIMO SE}) in the integration in (\ref{eq:SE avg gen}) we obtain, as detailed again in App. \ref{siso_avg},
\begin{align}\nonumber
\!\!\!\!\! \Cbar^{\scriptscriptstyle \mathsf{MIMO}}_{\scriptscriptstyle 2\times2} & \approx \frac{s^\star \, \log_2 e}{(s^\star-1)\,e^{\frac{s^\star-1}{D_\delta}}} + {}_1F_1 \big(\delta,1+\delta,-1 \big) \, \sinc( \delta)  \log_2 e\\ 
	\nonumber
	&\quad + 2.8 \Big[ {}_2F_1 \big(1,\delta,1+\delta,-2.44 \big) + \delta \log (1.41) \Big] \, \frac{\sinc (\delta)}{\delta} \\
	&\quad +2.8 \bigg[ \frac{E_1\left(-s^\star \,\left(0.41+1/D_\delta\right) \right)}{e^{0.41 \, s^\star}} - E_1\left(-s^\star/D_\delta\right) +  e^{s^\star/D_\delta} \, \log (1+0.41\,D_\delta) \bigg].
	\label{eq:MIMO avg SE3}
\end{align}

\begin{exmp}
\label{BIST2}
Fig. \ref{Fig:val3} compares (\ref{eq:MIMO avg SE1}) and (\ref{eq:MIMO avg SE3}) against $\barCexact$.
For $\eta=4$ in particular, (\ref{eq:MIMO avg SE3}) returns
\begin{align}
\Cbar^{\scriptscriptstyle \mathsf{MIMO}}_{\scriptscriptstyle 2\times2} &\approx 0.26 + \frac{\log_2 e}{\sqrt{\pi}} \erf(1) + \frac{5.6}{\pi} \left[\sqrt{0.41} \left(\pi - 2 \arctan(\sqrt{0.41}) \right) + \log 1.41 \right] \\
&= 3.84 \nonumber
\end{align}
while the Monte-Carlo average of its $\Cexact$ counterpart is $3.87$.

\end{exmp}

Combining Examples \ref{SECDF_valid3} and \ref{BIST2}, two-antenna single-user MIMO is seen from our analysis to provide a $93\%$ increase in the spectral efficiency of an entire interference-limited network, a determination that would classically have entailed very extensive simulations.

For higher-order MIMO, the integration of $C^{\scriptscriptstyle \mathsf{MIMO}}_{\scriptscriptstyle \Nr\times\Nt}(\rho)$ over $F_\rho(\cdot)$ in (\ref{eq:SIRCDF_asymptotic2}) yields the single-integral solution
%
\begin{align}
\nonumber
\! \!\bar{C}^{\scriptscriptstyle \mathsf{MIMO}}_{\scriptscriptstyle \Nr\times\Nt}\!\! &\approx  \log_2 e \! \sum_{i=0}^{m-1} \sum_{j=0}^{i} \sum_{\ell=0}^{2\,j} \Bigg\{\! \binom{2\,i-2\,j}{i-j}  \binom{2\,j+2\,n-2\,m}{2\,j-\ell} \frac{(-1)^{\ell} \, (2\,j)! \, (n-m+\ell)!}{2^{2i-\ell} \, j! \, \ell! \, (n-m+j)!}  \\
& \quad \quad \cdot \sum_{q=0}^{n-m+\ell} \int_{0}^{\infty} \frac{1}{(1+\gamma)^{q+1}} \left[ \frac{s^\star}{s^\star-\gamma\,\Nt} \,  e^{\frac{s^\star- \gamma \, \Nt}{D_\delta}} + \delta \, \sinc (\delta) \frac{\bar{\Gamma}(\delta, \gamma \, \Nt)}{(\gamma \, \Nt)^\delta}\right] {\rm d}\gamma \Bigg\}. \label{eq:higher order MIMO exp}
\end{align}
Although, for some antenna configurations beyond $\Nt=\Nr=2$, it may be possible to express this integral via special functions, it is beside the point once the expressions become overly intricate. Rather, it seems preferable to directly integrate numerically for each specific path loss exponent of interest.

\begin{table}
	\caption{Spatially averaged ergodic spectral efficiency $\bar{C}^{\scriptscriptstyle \mathsf{MIMO}}_{\scriptscriptstyle \Nr\times\Nt}$ with $\eta=4$.}
	\label{tab:MIMOavgseff eta4}
	\begin{center}
		\begin{tabular}{|c||c|c|c|c|}
			\hline
			\diagbox{$\Nt$}{$\Nr$} & $1$  & $2$ & $3$ & $4$ \\ \hline \hline
			$1$ & $1.99$  & $2.76$ & $3.25$ & $3.62$ \\ \hline   
			$2$ & $2.13$  & $3.84$ & $4.79$ & $5.48$ \\ \hline   
			$3$ & $2.18$  & $4.11$ & $5.71$ & $6.75$ \\ \hline   
			$4$ & $2.21$  & $4.24$ & $6.05$ & $7.59$ \\ \hline   
		\end{tabular}
	\end{center} 
\end{table}

\begin{exmp}
Shown in Table \ref{tab:MIMOavgseff eta4} is $\bar{C}^{\scriptscriptstyle \mathsf{MIMO}}_{\scriptscriptstyle \Nr\times\Nt}$ computed via (\ref{eq:higher order MIMO exp}) for  different values of $\Nt$ and $\Nr$ with the path loss exponent $\eta=4$.
\end{exmp}

\section{Sectorization}
\label{sectorization}

\begin{figure}
  \centering
  \subfloat[No sectorization ($S=1$).]{\includegraphics[width=0.35\linewidth]{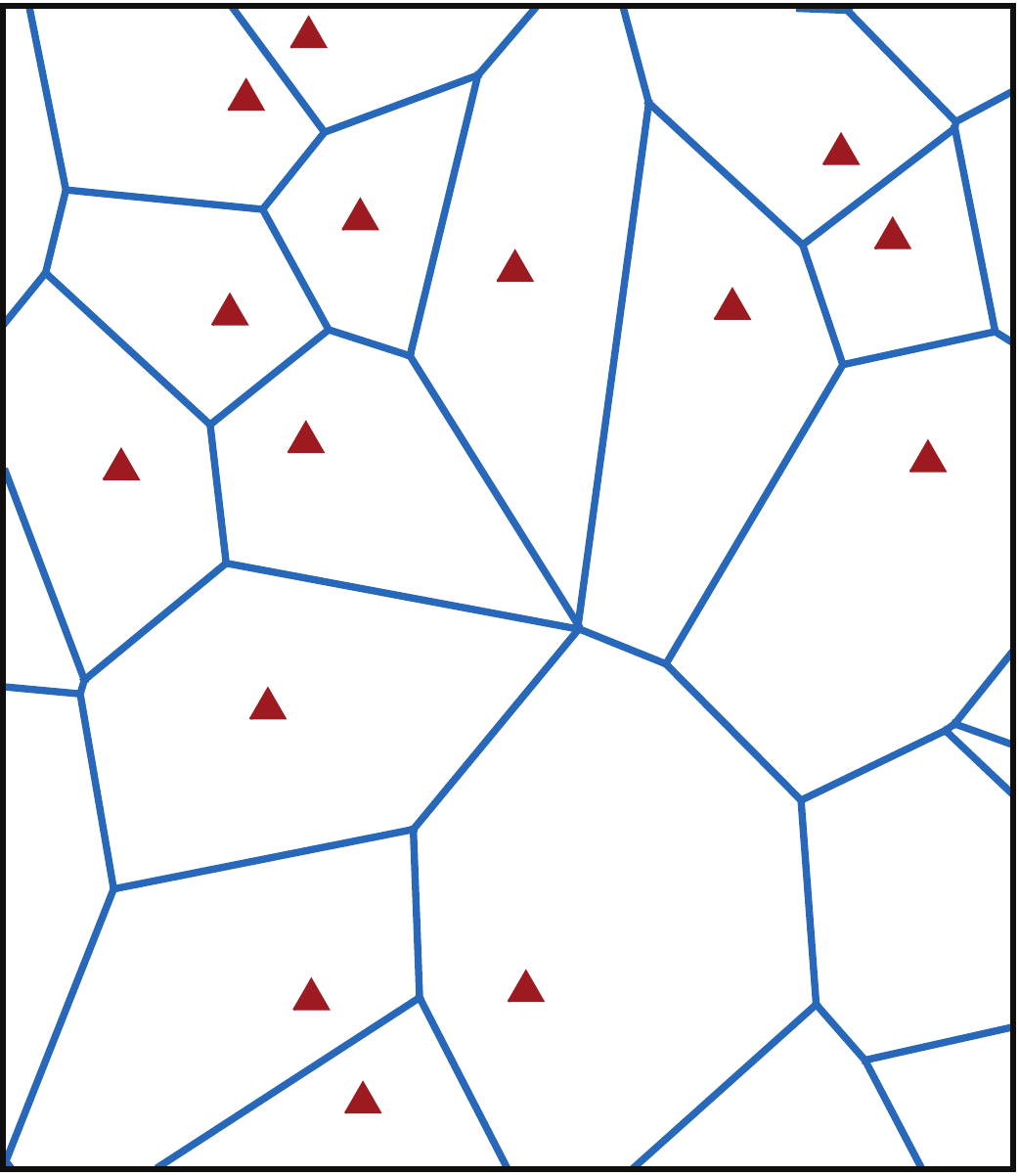}\label{fig:f1}}
\quad
  \subfloat[Tri-sectorized cells ($S=3$).]{\includegraphics[width=0.35\linewidth]{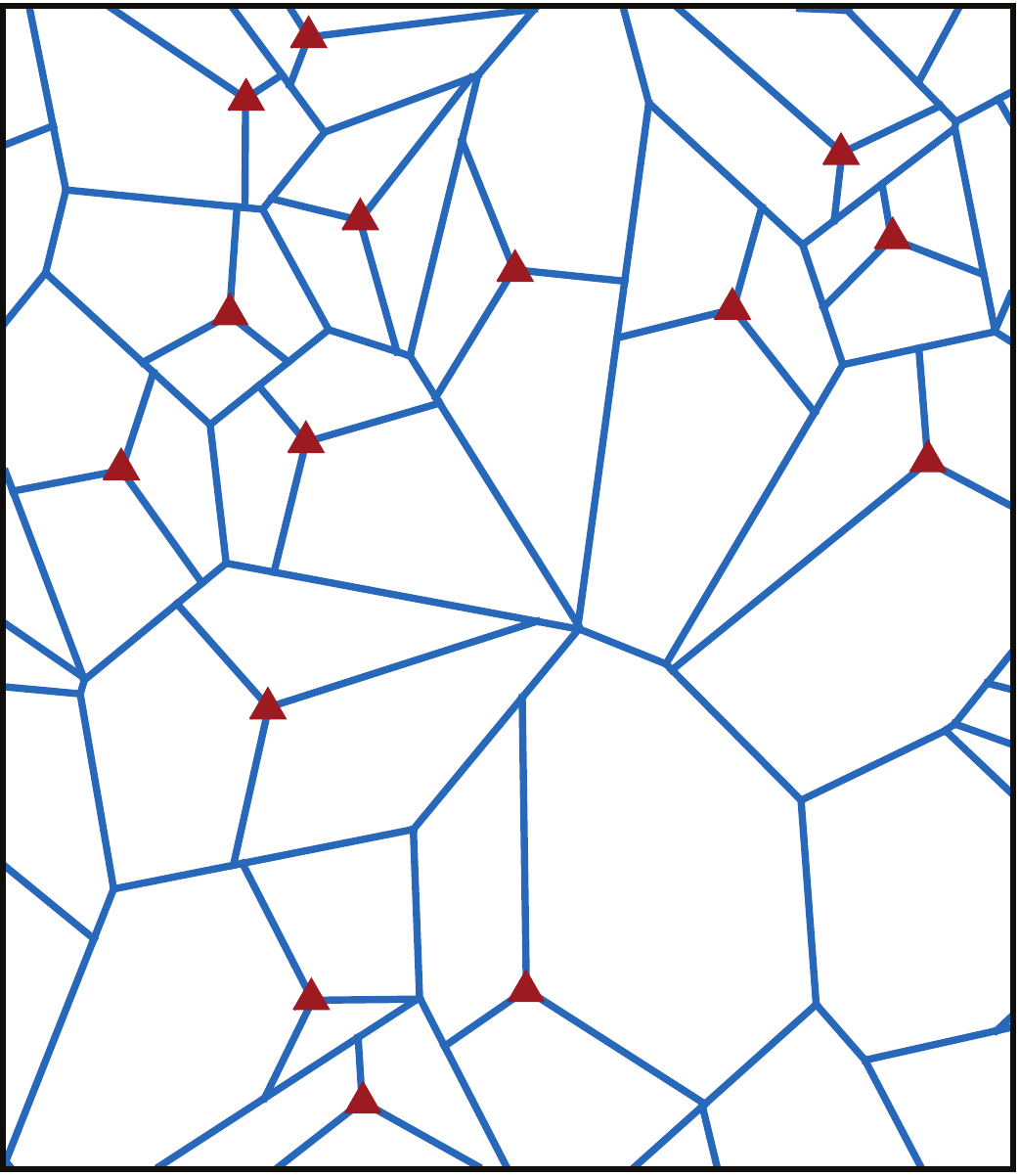}\label{fig:f2}}
  \caption{Tesselation of a Poisson network, with and without sectorization.}
  \label{Fig:sectors}
\end{figure}

Let us now incorporate cell sectorization to the model. Each BS, now allowed to comprise $S$ sector antennas uniformly staggered in azimuth, communicates with one user per channel and per sector (cf. Fig. \ref{Fig:sectors}).
From the vantage of the user at the origin, the downlink signals from the sectors of any given BS undergo the same path loss and shadowing but different antenna gains.

\begin{figure} 
	\centering
	\includegraphics [width=0.3\columnwidth]{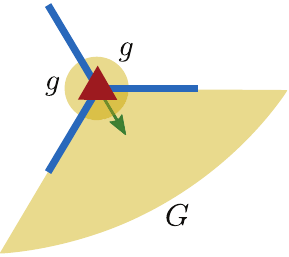}
	\caption{Antenna pattern for $S=3$ with in-sector gain $G$ and out-of-sector gain $g$.}
	\label{Fig:antenna_pat}
\end{figure}

Given an arbitrary azimuth pattern for the sector antennas, \cite{7243344} characterized the local-average SINR distribution in the Laplace domain.
Differently, in this paper, we provide direct expressions that rely only on the antenna front-to-back ratio $\Gfbr \geq 1$ and on the number of sectors, $S$. Given an in-sector gain $G = \frac{\Gfbr \, S}{\Gfbr+S-1}$ and an out-of-sector gain $g =\frac{S}{\Gfbr+S-1}$, our model for the antenna pattern as a function of the azimuth $\phi$ is (cf. Fig. \ref{Fig:antenna_pat})
\begin{align}\label{eq:ant_pat}
g_{\scriptscriptstyle S}(\phi) = \left\{ \begin{array}{l l}
			G \qquad\qquad &  \phi_0 - \pi/S \leq \phi < \phi_0 + \pi/S \\
			g  \qquad\qquad &   \mathrm{elsewhere}, 
						\end{array} \right.
\end{align}
where $\phi_0$ indicates the orientation of the antenna.
This way, it is ensured that the total radiated power is preserved, i.e., that
\begin{align}
\int\limits_{0}^{2 \pi} \frac{g_{\scriptscriptstyle S}(\phi)}{2  \pi} \, {\rm d}\phi = 1.
\end{align}
Setting $S=1$ we recover an unsectorized network where $g_{\scriptscriptstyle 1}(\phi) = 1$. In turn, for $\Gfbr \rightarrow \infty$, the $S$ sectors become ideal as $G \rightarrow S$ and $g \rightarrow 0$.
Under the foregoing model, the intended signal from the serving sector has gain $G$ while the $(S-1)$ interfering transmissions from other sectors of the same BS have gain $g$. The $S$ transmissions from every other BS add to the interference. 

%

The small-scale fading in the link from each sector is independent and of unit-power, with the receiver knowing only the fading experienced by the intended signal.

\subsection{Local-Average SINR}

With $P$ now denoting the per-sector transmit power, the total power that each interfering BS launches towards the user at the origin is  $ P \big( G+(S-1)\,g \big) = P S$. Since the useful signal launched by the intended BS is $P G$, the local-average SINR at the origin is
\begin{align}
\label{eq:localavgSIR_sector}
\rho_{\scriptscriptstyle S} 
&=\frac{P \, G \, \ro^{-\eta}}{P \, (S-1) \, g \, \ro^{-\eta} + P \, S \sum_{k=1}^{\infty} \rk^{-\eta} + N_0}
\end{align}
irrespective of how the sectors are oriented at each BS. (This orientation-invariance is not an artifact of our model; rather, it has been shown that the distribution of $\rho_{\scriptscriptstyle S}$ is insensitive to the sector orientations regardless of the
antenna patterns  \cite{7243344}.)
%

In interference-limited conditions, the local-average SINR becomes
\begin{align}
\rho_{\scriptscriptstyle S} &=\frac{\Gfbr \, \ro^{-\eta}}{(S-1) \, \ro^{-\eta} + (\Gfbr + S-1)  \sum_{k=1}^{\infty} \rk^{-\eta} } ,
\label{CUPsucks}
\end{align}
which, with ideal sectorization, i.e., for $Q \rightarrow \infty$, converges to its unsectorized self, $\rho = \ro^{-\eta} / \sum_{k=1}^{\infty} \rk^{-\eta}$.
It follows that, under ideal sectorization, all the results derived for unsectorized networks continue to apply, only on a per-sector rather than a per-BS basis.
Conversely, under nonideal sectors, the CDF of $\rho_{\scriptscriptstyle S}$ can be obtained as
\begin{align}\nonumber
\!\!\!\!\!F_{\rho_{\scriptscriptstyle S}}(\theta) &= \mathbP\!\left[\!\frac{\Gfbr \, \ro^{-\eta}}{(S-1) \, \ro^{-\eta} + (\Gfbr + S-1)  \sum_{k=1}^{\infty} \rk^{-\eta} } < \theta \right] \\
&= \mathbP\left[\frac{\ro^{-\eta}}{\sum_{k=1}^{\infty} \rk^{-\eta}} < \frac{\Gfbr+S-1}{\Gfbr/\theta-S+1}\right]\\ 
	&= \left\{ \begin{array}{l l}
	F_{\rho}\left[\frac{\Gfbr+S-1}{\Gfbr/\theta-S+1}\right] \qquad\qquad &  0 \leq \theta < \frac{\Gfbr}{S-1} \\
	1 \qquad\qquad &     \theta \geq \frac{\Gfbr}{S-1}  ,
	\end{array} \right. \label{eq:CDF_SIR_sector1}
\end{align}
which is capped at $\frac{Q}{S-1}$ because of interference among same-BS sectors.

\begin{figure}
	\centering
	\includegraphics [width=0.75\columnwidth]{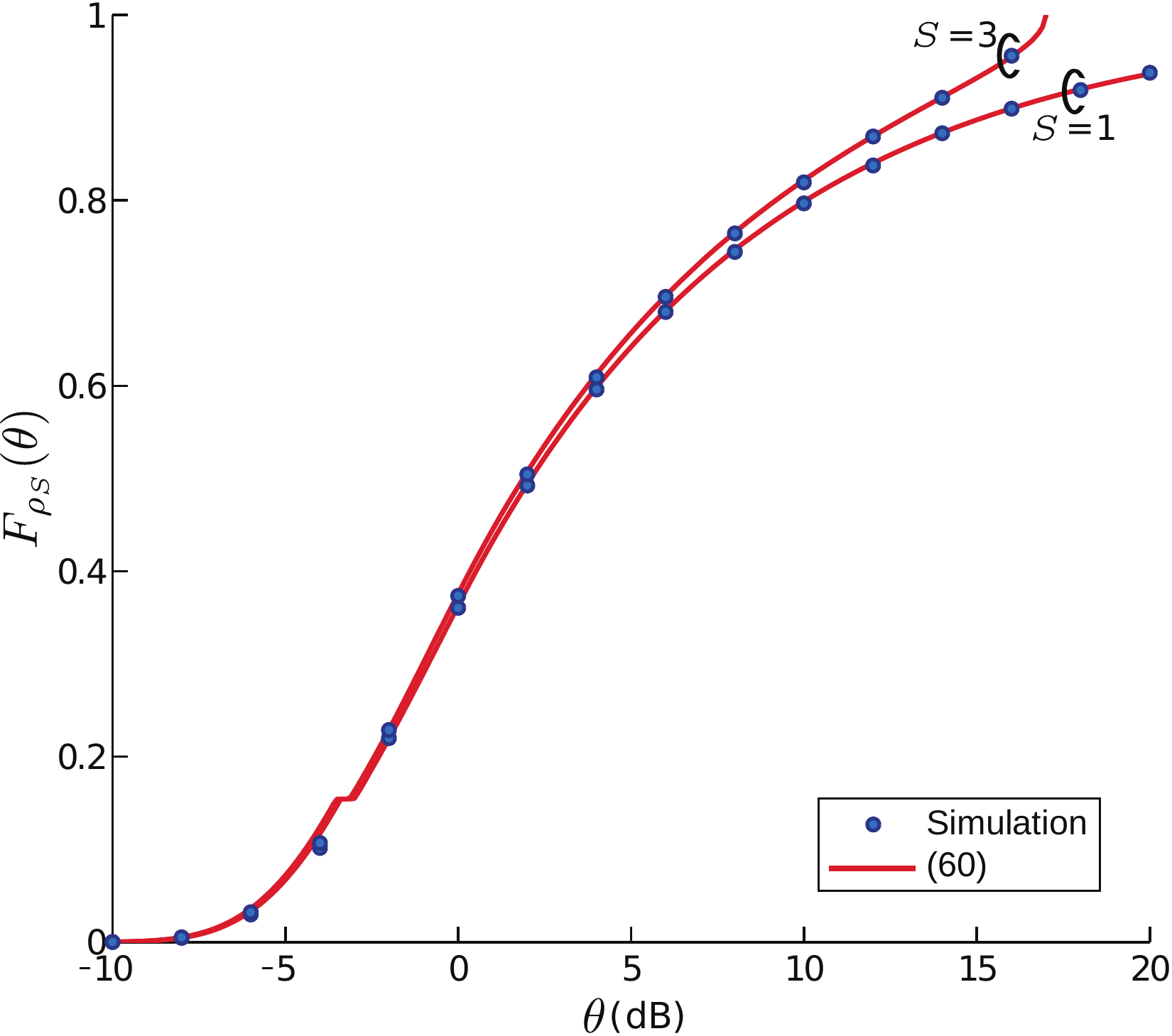}
	\caption{$F_\rho(\cdot)$ without sectorization and $F_{\rho_3}(\cdot)$ with $Q=20$ dB, both for $\eta=4$.}
	\label{RhoSECT}
\end{figure}

Plugging (\ref{eq:SIRCDF_asymptotic1})  into (\ref{eq:CDF_SIR_sector1}) gives 
\begin{align}
\label{eq:SIRCDF_sector}
 \left\{ \begin{array}{l l}
 \!\! F_{\rho_{\scriptscriptstyle S}}(\theta)  \simeq	e^{s^\star \frac{\Gfbr/\theta-S+1}{\Gfbr+S-1} } \qquad\qquad &  0 \leq \theta < \frac{s^\star \, \Gfbr/\log A_\delta}{\Gfbr \, + \,(1\,+\,s^\star/\log A_\delta) (S-1)} \\
 \!\!F_{\rho_{\scriptscriptstyle S}}(\theta)  \approx	A_\delta  \qquad\qquad &   \frac{s^\star \, \Gfbr /\log A_\delta}{\Gfbr \, + \,(1\,+\,s^\star/\log A_\delta) (S-1)} \leq \theta < \frac{\Gfbr}{2\,\Gfbr \,+ \, 3\,(S-1)} \\
 \!\! F_{\rho_{\scriptscriptstyle S}}(\theta)  =	1 -  \frac{\sinc \, \delta}{\left(\frac{\Gfbr+S-1}{\Gfbr/\theta-S+1}\right)^{\delta}} + B_{\delta} \! \left[\frac{\Gfbr+S-1}{\Gfbr\,(1/\theta-1)-2\,(S-1)}\right]  \qquad &  \frac{\Gfbr}{2\,\Gfbr \,+ \, 3\,(S-1)} \leq \theta < \frac{\Gfbr}{\Gfbr \,+ \, 2\,(S-1)} \\
 \!\! F_{\rho_{\scriptscriptstyle S}}(\theta)  =	1 - \frac{\sinc \, \delta}{\left(\frac{\Gfbr+S-1}{\Gfbr/\theta-S+1}\right)^{\delta}}  \qquad\qquad &   \frac{\Gfbr}{\Gfbr \,+ \, 2\,(S-1)} \leq \theta < \frac{\Gfbr}{S-1} \\
\!\! F_{\rho_{\scriptscriptstyle S}}(\theta)  =	1 \qquad\qquad &  \theta \geq \frac{\Gfbr}{S-1} ,
 \end{array} \right.
\end{align}
which, for $\eta = 4$, specializes to
\begin{align}
\left\{ \begin{array}{l l}
 \!\! F_{\rho_{\scriptscriptstyle S}}(\theta) \simeq e^{-0.854 \frac{\Gfbr/\theta - S+1}{\Gfbr+S-1} } \qquad\quad &  0 \leq \theta < \frac{0.457 \, \Gfbr}{\Gfbr + 1.457 \, (S-1)} \\
 \!\! F_{\rho_{\scriptscriptstyle S}}(\theta) \approx 0.154 \qquad\quad &    \frac{0.457 \, \Gfbr}{\Gfbr + 1.457 \, (S-1)} \leq \theta <  \frac{\Gfbr}{2\,\Gfbr \,+ \, 3\,(S-1)} \\
 \!\! F_{\rho_{\scriptscriptstyle S}}(\theta) = 1 - \frac{4}{\pi}  \sqrt{\frac{\Gfbr/\theta - S+1}{\Gfbr+S-1}} +  \frac{\Gfbr}{\pi}  \frac{1+1/\theta}{\Gfbr+S-1} \qquad\quad &    \frac{\Gfbr}{2\,\Gfbr \,+ \, 3\,(S-1)} \leq \theta < \frac{\Gfbr}{\Gfbr \,+ \, 2\,(S-1)} \\
 \!\! F_{\rho_{\scriptscriptstyle S}}(\theta) = 1 - \frac{2}{\pi}  \sqrt{\frac{\Gfbr/\theta - S+1}{\Gfbr+S-1}} \qquad\quad &  \frac{\Gfbr}{\Gfbr \,+ \, 2\,(S-1)} \leq \theta < \frac{\Gfbr}{S-1}  \\
 \!\! F_{\rho_{\scriptscriptstyle S}}(\theta) = 1 \qquad\quad &  \theta \geq \frac{\Gfbr}{S-1}.
 \end{array} \right. \label{eq:CDF_SIR_sector2}
\end{align}
Alternatively, plugging (\ref{eq:SIRCDF_asymptotic2}) into (\ref{eq:CDF_SIR_sector1}), a simpler and slightly less accurate expression is obtained 
\begin{align}
\label{eq:SIRCDF_asymptotic3}
 \left\{ \begin{array}{l l}
\!\! F_{\rho_{\scriptscriptstyle S}}(\theta) \simeq e^{s^\star \frac{\Gfbr/\theta-S+1}{\Gfbr+S-1}} \qquad\quad &  0 \leq \theta < \frac{s^\star \, \Gfbr/\log (1-\sinc \, \delta)}{\Gfbr \, + \,[1\,+\,s^\star/\log  (1-\sinc \, \delta)] (S-1)}  \\
\!\! F_{\rho_{\scriptscriptstyle S}}(\theta) \approx 1 -\sinc \, \delta  \qquad\quad &    \frac{s^\star \, \Gfbr/\log (1-\sinc \, \delta)}{\Gfbr \, + \,[1\,+\,s^\star/\log  (1-\sinc \, \delta)] (S-1)} \leq \theta < \frac{\Gfbr}{\Gfbr \,+ \, 2\,(S-1)} \\
\!\! F_{\rho_{\scriptscriptstyle S}}(\theta) = 1 - \frac{\sinc \, \delta}{\left(\frac{\Gfbr+S-1}{\Gfbr/\theta-S+1}\right)^{\delta}} \qquad\quad &  \frac{\Gfbr}{\Gfbr \,+ \, 2\,(S-1)}  \leq  \theta <  \frac{\Gfbr}{S-1} \\
\!\! F_{\rho_{\scriptscriptstyle S}}(\theta) = 1 \qquad\quad &  \theta \geq  \frac{\Gfbr}{S-1} .
\end{array} \right.
\end{align}


\begin{exmp}
Let $\eta=4$. Shown in Fig. \ref{RhoSECT} are $F_\rho(\cdot)$ without sectorization and $F_{\rho_3}(\cdot)$ with $Q=$ 20 dB, which is a rather typical front-to-back ratio.
\end{exmp}

Although, as Fig. \ref{RhoSECT} visualizes, nonideal sectorization puts a hard ceiling on the SIR, this only affects the distribution modestly. In exchange, the available bandwidth gets to be reused $S$ times per cell and thus sectorization is decidedly advantageous. 

\subsection{Ergodic Spectral Efficiency}

As in Section \ref{SectionSE1}, the CDF of the ergodic spectral efficiency $F_C(\cdot)$ is determined by mapping the applicable function $C(\rho_{\scriptscriptstyle S})$ onto $F_{\rho_{\scriptscriptstyle S}}(\cdot)$. 
For Rayleigh-faded SISO channels in particular, we can invoke (\ref{eq:Erg Seff CDF approx}) to explicitly express this mapping as
\begin{align}\label{eq:CDF_SE_siso_sect}
F_C(\gamma) \approx	F_{\rho_{\scriptscriptstyle S}}\left(  \frac{e^{\frac{\gamma}{1.4}} -1 }{0.82} \right).
\end{align}
\begin{exmp}
		\begin{figure}
			\centering
			\includegraphics [width=0.75\columnwidth]{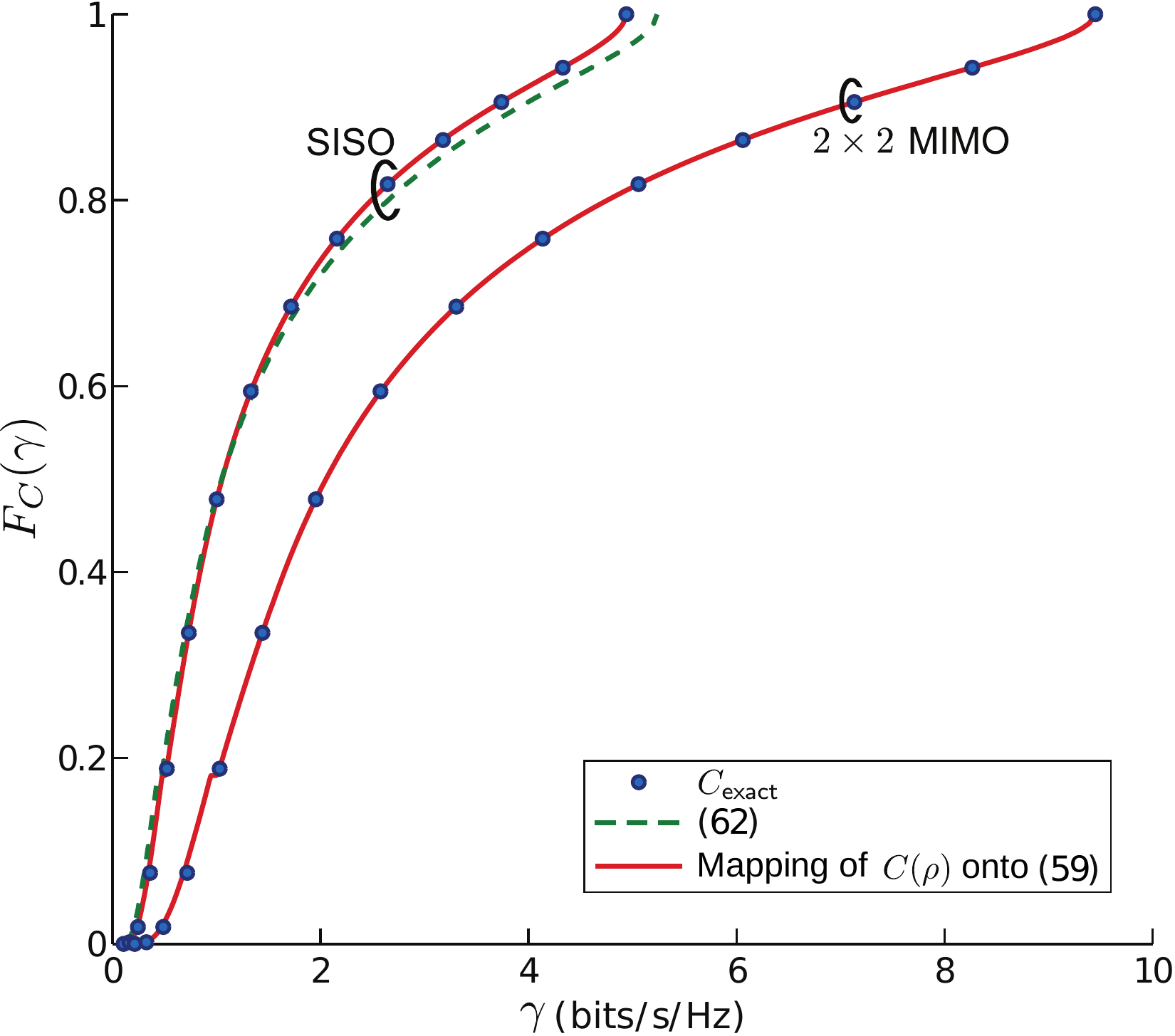}
			\caption{CDF of ergodic spectral efficiency for $\eta=3.8$, $S=3$ and $\Gfbr= 20$ dB.}
			\label{Fig:val4}
		\end{figure}
	\label{SECDF_valid4}
	Reconsider Examples \ref{SECDF_valid} and \ref{SECDF_valid2}, but now with $S=3$ sectors having front-to-back ratio $\Gfbr = 20$ dB. 
	Shown in Fig. \ref{Fig:val4} are (\ref{eq:CDF_SE_siso_sect}), the numerical mapping of $C(\rho)$ onto (\ref{eq:SIRCDF_sector}) for SISO and MIMO, and the simulated $\Cexact$.
\end{exmp}

For any setting for which $C(\rho_{\scriptscriptstyle S})$ is available, we can also compute the spatially averaged ergodic spectral efficiency as
\begin{align}\label{eq:SE avg gen2}
\Cbar &= \int_{0}^{{\frac{\Gfbr}{S-1}}} C(\theta) \, {\rm d} F_{\rho_{\scriptscriptstyle S}} (\theta) .
\end{align}

\begin{exmp}
Let $\eta=3.8$, $S=3$ and $Q=20$ dB. For SISO, $\Cbar = 1.53$ b/s/Hz per sector, which adds up to $4.59$ b/s/Hz per BS. 
For $S=1$, in contrast, a read-out of Fig. \ref{Fig:val3} gives $\Cbar=1.84$ b/s/Hz per BS.
With the sector nonideality accounted for, the overall spectral efficiency of a SISO network gets multiplied by $2.5$ when cells are split in three sectors.

For $2 \times 2$ MIMO, in turn, 
$\Cbar = 2.95$ b/s/Hz per sector adding up to $8.85$ b/s/Hz per BS. 
The sectorization multiplier is $2.49$, almost unchanged from its SISO value.


\end{exmp}

\section{Application to Lattice Networks}
\label{hexagons}

Before wrapping up, we close the loop and verify the initial premise whereby a PPP model was invoked for the BS locations, confirming that such model is representative because of shadowing. To that end, we compare our PPP-based analytical results with Monte-Carlo simulations for settings at the extreme end of the point process spectrum, namely regular lattices.
\begin{figure} 
	\centering
	\includegraphics [width=0.75\columnwidth]{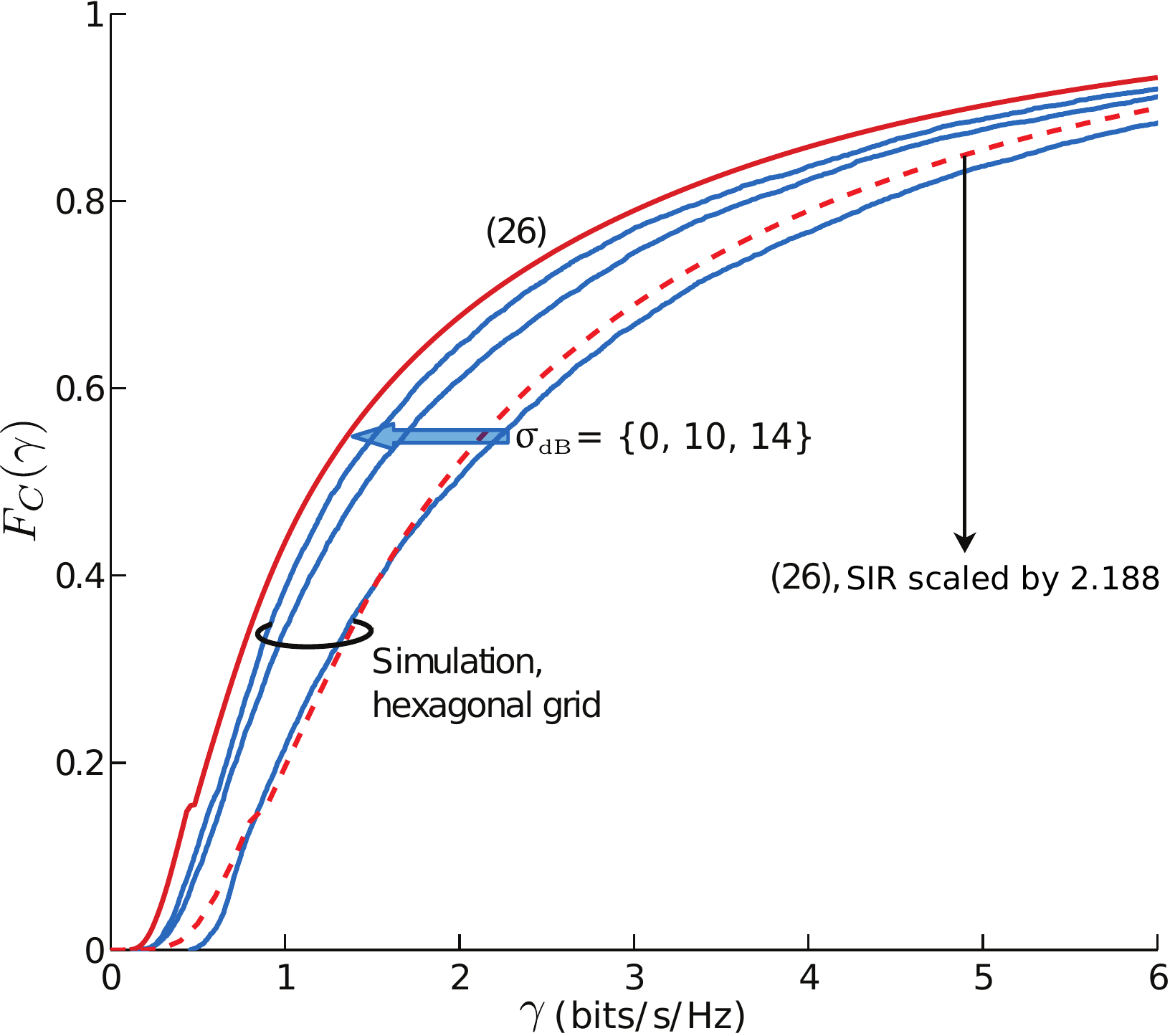}
	\caption{CDF of SISO ergodic spectral efficiency with no sectorization ($S=1$): PPP-based analytical result vs Monte-Carlo over a lattice network of hexagonal cells with $\shadSD=0$ dB, $10$ dB and $14$ dB. The dashed curve corresponds to using $F_\rho(\theta/2.188)$ in lieu of $F_\rho(\theta)$ 
		and proceeding as if the network conformed to a PPP. 
		In all cases, $\eta = 4$.}
	\label{Fig:hex1}
\end{figure}
\begin{exmp}
\label{VagaMetro}
Shown in Fig. \ref{Fig:hex1} is $F_C(\cdot)$ as given in (\ref{esterilla}) for SISO, in comparison with the results for a lattice of $977$ hexagonal cells with $\eta=4$ and various values of $\shadSD$.
The convergence is conspicuous and, most importantly, the agreement is excellent for typical outdoor values of $\shadSD$, in the range of $10$--$12$ dB. 
\end{exmp}

		\begin{figure} 
			\centering
			\includegraphics [width=0.75\columnwidth]{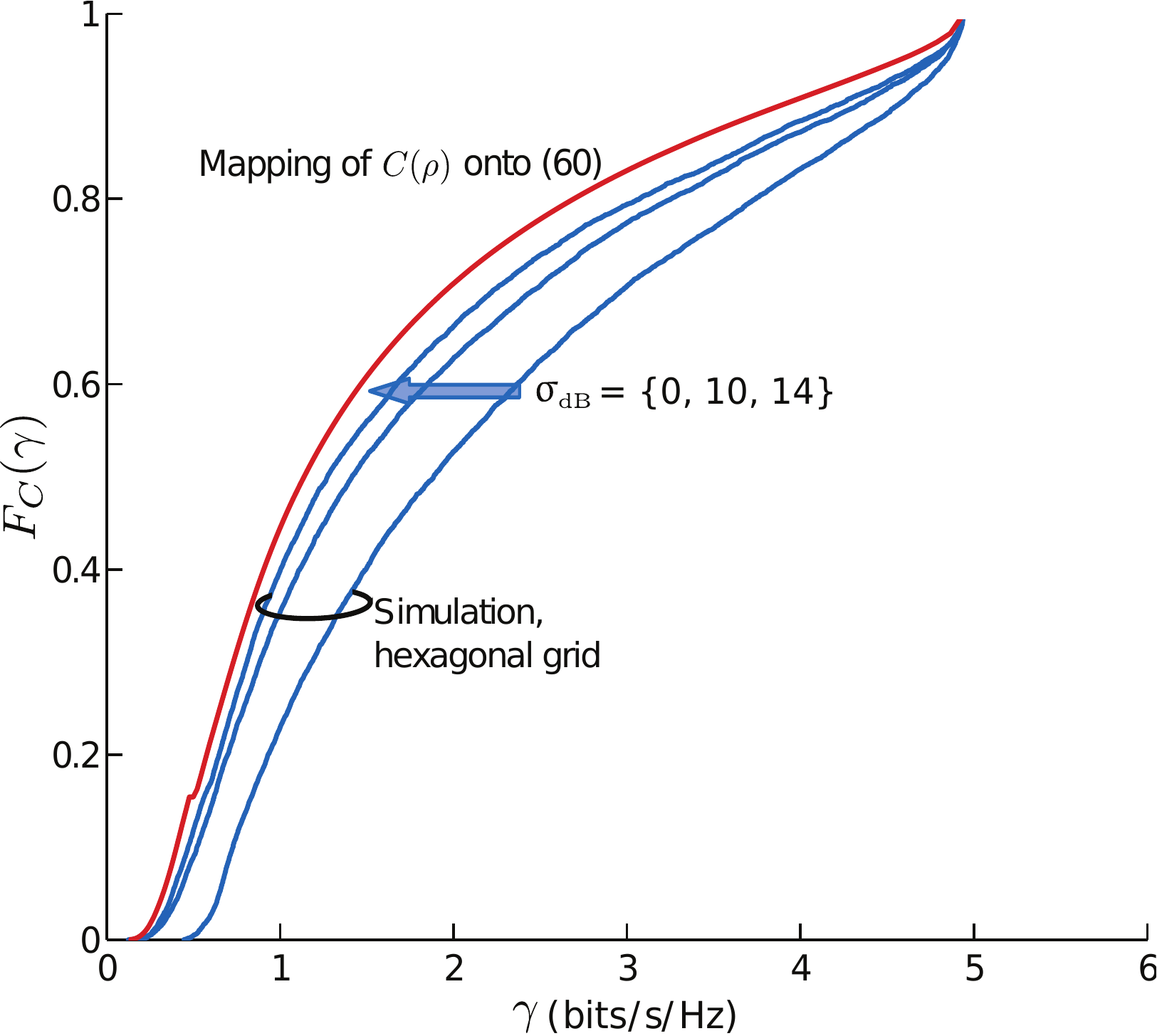}
			\caption{CDF of SISO ergodic spectral efficiency with $S=3$ sectors and $\Gfbr = 20$ dB: PPP-based analytical result vs Monte-Carlo over a lattice network of hexagonal cells with $\shadSD=0$ dB, $10$ dB and $14$ dB. In all cases, $\eta = 4$.}
			\label{Fig:hex2}
		\end{figure}

\begin{exmp}
The comparison of Fig. \ref{Fig:hex1} is repeated, for $S=3$ sectors and $\Gfbr = 20$ dB, in  Fig. \ref{Fig:hex2}, with similar observations in terms of the convergence.
\end{exmp}

\begin{figure} 
	\centering
	\includegraphics [width=0.75\columnwidth]{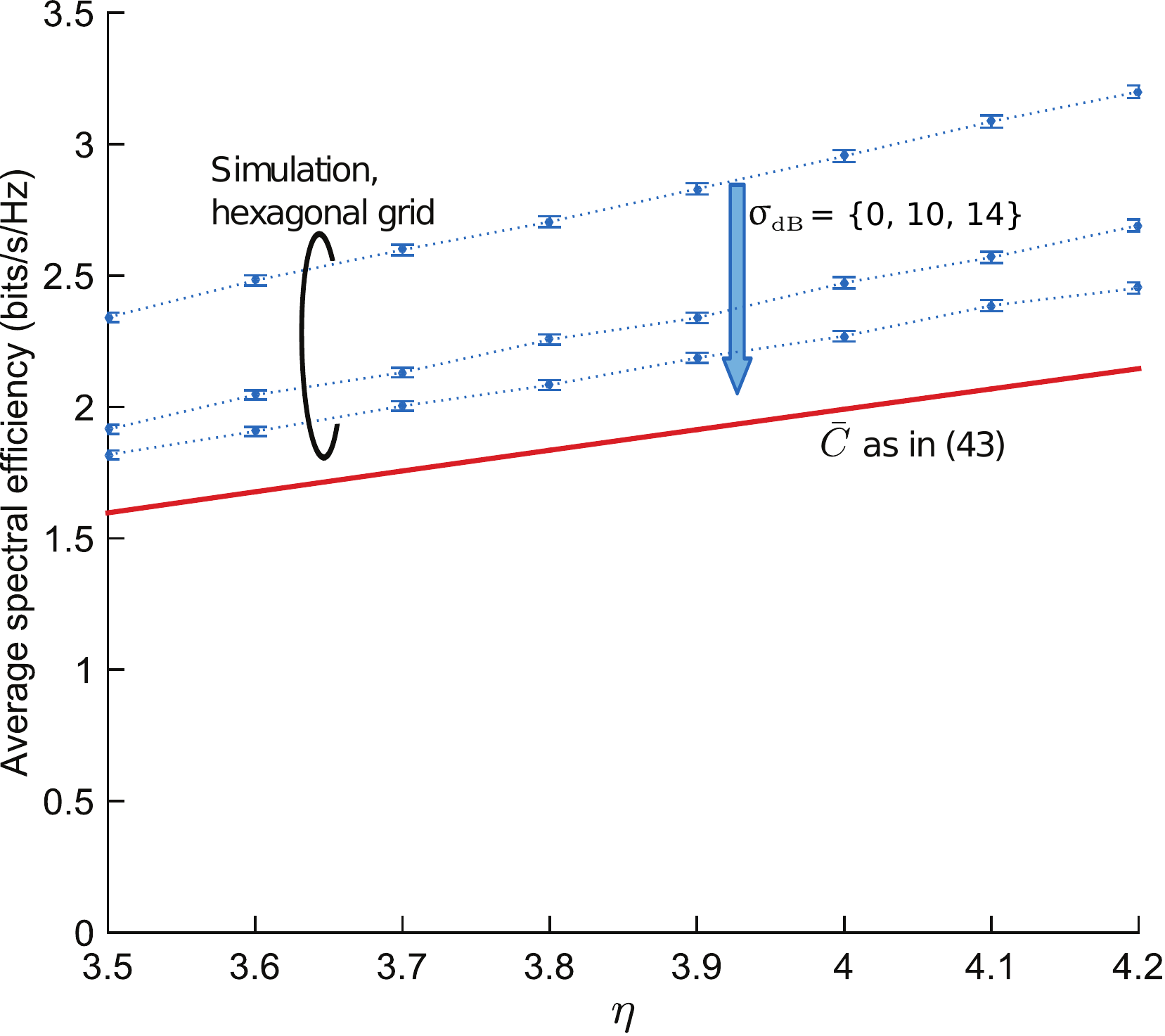}
	\caption{Spatially averaged SISO ergodic spectral efficiency as function of $\eta$ with no sectorization ($S=1$): PPP-based analytical result vs Monte-Carlo on a lattice network of hexagonal cells with $\shadSD=0$ dB, $10$ dB and $14$ dB. The error bars around the simulation results indicate the $99\%$ confidence interval.}
	\label{Fig:avg_SE_hex1}
\end{figure}

\begin{exmp}
\label{AraBus}

	Shown in Fig. \ref{Fig:avg_SE_hex1} is $\Cbar$ as given in (\ref{eq:Avg seff asymp}), for SISO, in comparison with the results for a lattice of $977$ hexagonal cells with various values of $\shadSD$ and $\eta$. 
	
\end{exmp}

%
%
%

Further reinforcing the relevance of PPP-based results to lattice networks, it has been argued in \cite{net:Guo14icc,net:Haenggi14wcl,net:Ganti16twc} that the SINR distribution of a shadowless lattice network is essentially a shifted version of its PPP counterpart, i.e., a shifted version of the SINR distribution for asymptotically strong shadowing. Moreover, this shift does not depend on the path loss exponent, but only on the type of lattice.

\begin{exmp}
For a triangular lattice (hexagonal cells), the shift to apply is $10 \log_{10} 2.188 = 3.4$ dB  \cite{net:Guo14icc,net:Haenggi14wcl,net:Ganti16twc}.
Shown in Fig. \ref{Fig:hex1} is $F_C(\cdot)$ recomputed with $F_\rho(\theta/2.188)$ in lieu of $F_\rho(\theta)$, and the agreement with $F_C(\cdot)$ for a network of hexagonal cells and no shadowing is highly satisfactory.
\end{exmp}

Altogether then, PPP results enable characterizing the distributions of SINR and spectral efficiency both with (asymptotically strong) shadowing and without, bracketing what a given network can exhibit over all possible shadowing strengths.


\section{Summary}

By decoupling small- and large-scale channel features and abstracting the former via local ergodicity, the stochastic geometry analysis of wireless networks can focus crisply on the large-scale properties. Jointly with a Gaussian model for the aggregate interference that recognizes that the fading of each term therein is unknown, this enables circumventing analytical roadblocks and deriving expressions that are more open to generalization, readily accommodating aspects such as MIMO or sectorization. The obtained spectral efficiencies lower-bound the exact values with a degree of accuracy that justifies writing $C \lesssim \Cexact$.

Thanks to the PPP-likeness brought about by shadowing, the obtained expressions apply to any stationary and ergodic network model exhibiting reasonable shadowing strengths. Furthermore, with a proper shift, these expressions apply to shadowless networks, which are relevant insofar as they can model planned deployments where the BS locations do depend on the radio propagation. Thus, PPP analysis can serve to bracket the entire performance range in a given environment.


The approach propounded in this paper is extensible to settings where noise is nonnegligible, and the accuracy of the results could then only improve further since our interference model is an exact match for Gaussian noise devoid of fading.

Besides the addition of noise, numerous other extensions are invited, for instance multiuser MIMO or BS cooperation \cite{lozano2013fundamental}. When dealing with MIMO, care must be exercised whenever $\Nr > \Nt$ and, especially, whenever $\Nr \gg \Nt$; then, if the fading of dominant interferer(s) can be tracked, spatial color can be exploited \cite{lozano2002capacity,Lee-Lee-Baccelli-16}. This can be accounted for by circumscribing our unfaded interference model to the rest of the interference, separately incorporating the terms that correspond to interferers whose fading is known.


\section*{Acknowledgment}

Motivating discussions with Prof. Jeffrey G. Andrews are gratefully acknowledged.
The efficient editorial handling by Dr. Bruno Clerckx and the excellent feedback provided by the reviewers are also gratefully acknowledged.

\appendices
\section{Computation of $\Cexact$}
\label{App:prelims}

For given $\{\rk\}_{k\in\mathbb{N}_0}$ and $\Ho$, the exact mutual information under the non-Gaussian $z$ in (\ref{eq:aggr_z}) is 
\begin{align}
\nonumber
& \! \! \! I \! \left( \! s_0 ; \sqrt{P \, \ro^{-\eta}} \, \Ho \, \so + z  \!\right) \\
&  \! \! \!= \mathfrak{h} \! \left( \!\sqrt{P \, \ro^{-\eta}} \, \Ho \, \so + z  \!\right) \! - \mathfrak{h} \! \left( \!\sqrt{P \, \ro^{-\eta}} \, \Ho \, \so + z \, \Big| \so \right) \\
& \! \! \! = \mathfrak{h} \! \left(\sqrt{P \, \ro^{-\eta}} \, \Ho \, \so + z  \right)  \!- \mathfrak{h} \! \left( z \right)
\label{RogueOne}
\end{align}
where we have introduced the differential entropy 
$\mathfrak{h}(x) =  - \E \left[ \log f_{\boldsymbol{x}'}( \boldsymbol{x}' ) \right]$
with $\boldsymbol{x}' = [\Re(x) \,\, \Im(x)]^{\rm T}$. The expectations are
evaluated
via Monte-Carlo 
over the random variables $\{\sk\}_{k\in\mathbb{N}_0}$ and $\{\Hk\}_{k\in\mathbb{N}}$, and averaged over multiple realizations of $\Ho$ to obtain $\Cexact$ in (\ref{marathon}).
The large-scale distribution of $\Cexact$, and its average $\barCexact$, are obtained via multiple realizations of $\{\rk\}_{k\in\mathbb{N}_0}$. 

With MIMO, the foregoing computation involves
 channel matrices and signal vectors, with the differential entropy of a vector
  $\boldsymbol{x} \in \mathbb{C}^n$ being $\mathfrak{h}(\boldsymbol{x}) =  - \E \left[ \log f_{\boldsymbol{x}''}( \boldsymbol{x}'' ) \right]  $ with $\boldsymbol{x}'' = [\Re(\boldsymbol{x}^{\rm T}) \,\, \Im(\boldsymbol{x}^{\rm T})]^{\rm T}$.

\section{Derivation of (\ref{eq:SE_wrong})}
\label{clasico}

With SINR given by (\ref{eq:SIR_wrong}), 
\begin{align}\nonumber
& \mathbE  \big[ \log_2 \! \big(1+ \SINRo \, | \{\rk\}_{k=0}^\infty \big) \big]  \\
& \qquad \quad= \int_{0}^{\infty} \mathbP \big[ \log_2 \! \big(1+ \SINRo \, | \{\rk\}_{k=0}^\infty \big) > \upnu \big] \, {\rm d}\upnu\\
\label{eq:nonerg_derivation1}
 &\qquad \quad = \int_{0}^{\infty} \frac{\log_2 e}{1+x} \,  F^{\mathrm{c}}_{ \SINRo|\{\rk\}} (x)  \, {\rm d}x
\end{align}
where (\ref{eq:nonerg_derivation1}) follows from the variable change $y = \log_2(1+x)$ and the CCDF $F^{\mathrm{c}}_{ \SINRo|\{\rk\}} (\cdot)$ can be computed as
\begin{align}\nonumber
\!\!\!F^{\mathrm{c}}_{ \SINRo|\{\rk\}} (x) &= \mathbP \! \left[\frac{P \, \ro^{-\eta} |\Ho|^2 }{P \sum_{k=1}^{\infty} \rk^{-\eta} |\Hk|^2 + N_0} > x \, \bigg| \{\rk\}_{k=0}^\infty \right] \\
\label{eq:nonerg_derivation2}
&\!\!\!\!\!\!\!\!\!\!\!\! = \mathbE \! \left[e^{-x \, \ro^\eta \left( \sum_{k=1}^{\infty} \rk^{-\eta} |\Hk|^2 + N_0/P \right)} \, \Bigg| \{\rk\}_{k=0}^\infty \right] \\
&\!\!\!\!\!\!\!\!\!\!\!\!\!= e^{-x \, \ro^\eta \, N_0/P} \; \mathbE \! \left[ \prod_{k=1}^{\infty} e^{ -x \, \ro^\eta \rk^{-\eta} |\Hk|^2 } \, \Bigg| \{\rk\}_{k=0}^\infty \right] \\
\label{eq:nonerg_derivation3}
&\!\!\!\!\!\!\!\!\!\!\!\!\!= e^{-x \, \ro^\eta \, N_0/P} \prod_{k=1}^{\infty} \frac{1}{1+ x \left(\ro/\rk\right)^\eta}
\end{align}
where (\ref{eq:nonerg_derivation2}) follows from the exponential distribution of $|\Ho|^2$ and the expectation is over $\{ H_k \}_{k=1}^\infty$. In turn, (\ref{eq:nonerg_derivation3}) follows from the fact that $\{ H_k \}_{k=1}^\infty$ are IID.

\section{Derivations of (\ref{eq:Avg seff asymp}), (\ref{eq:MIMO avg SE1}) and (\ref{eq:MIMO avg SE3})}
\label{siso_avg}

Plugging the PDF obtained by differentiating (\ref{eq:SIRCDF_asymptotic2}) into (\ref{eq:SE avg gen}),
\begin{align}\label{eq:derivation1}
\Cbar \approx \int_{0}^{\frac{s^\star}{\log (1 -\sinc \, \delta)}} C(\theta) \, \frac{- e^{s^\star/\theta} \, s^\star}{\theta^2} \, {\rm d}\theta+  \int_{1}^{\infty} C(\theta) \, \frac{\delta \, \sinc\,\delta}{\theta^{\delta+1}} \, {\rm d}\theta.
\end{align}
From $C(\cdot)$ as given in (\ref{eq:SE_spec_geo}) and (\ref{eq:MIMO SE}), the above integrals yield  (\ref{eq:Avg seff asymp}) and (\ref{eq:MIMO avg SE1}), respectively. These integrations are facilitated by invoking
$
E_1(x)=-\text{E}_\text{i}(-x) 
$,
where $\text{E}_\text{i}(x) = \int_{-x}^{\infty}\frac{-e^{-t}}{t} \, {\rm d}t$, in conjunction with the identities given in~\cite{WOLFRAM1} with appropriate variable changes.

Similarly, from $C(\cdot)$ as given in  (\ref{eq:MIMOSE_approx}), integration by parts in (\ref{eq:derivation1}) using the identities \cite[2.325.6]{Gradshteyn-07}, \cite[2.325.7]{Gradshteyn-07}, \cite[2.728.1]{Gradshteyn-07} and \cite[3.194.2]{Gradshteyn-07} with appropriate variable changes gives the expression claimed in (\ref{eq:MIMO avg SE3}). 

%

\bibliographystyle{IEEEtran}
\bibliography{jour_short,conf_short,references}

\end{document}